\documentclass[11pt]{article}

\usepackage[utf8]{inputenc}
\usepackage[T1]{fontenc}
\usepackage{amsmath}
\usepackage{amsfonts}
\usepackage{amssymb}
\usepackage[version=4]{mhchem}
\usepackage{stmaryrd}
\usepackage{graphicx}
\usepackage{array, xcolor, lipsum, bibentry, fancyhdr}
\usepackage[export]{adjustbox}
\graphicspath{ {images/} }
\usepackage{hyperref}

\hypersetup{
    colorlinks,
    linkcolor={red!50!red},
    citecolor={blue!50!blue},
    urlcolor={blue!80!blue}
}

\newcounter{susis}

\newcommand{\myhyperlink}[2]{%
\refstepcounter{susis}\label{susislink#1}%
\ifcsname r@susistarget#1\endcsname
\hyperlink{#1}{#2}%
\else
{\textit{#2}}%
\fi
}

\newcommand{\myhypertarget}[2]{%
\refstepcounter{susis}\label{susistarget#1}%
\ifcsname r@susislink#1\endcsname
\hypertarget{#1}{#2}%
\else
\hypertarget{#1}{#2}%
\fi
}

\title{{ \bf MLP, XGBoost, KAN, TDNN, and LSTM-GRU Hybrid RNN with Attention for SPX and NDX European Call Option Pricing}}

\author{
  {\it Boris Ter-Avanesov$^{1}$ and Homayoon Beigi$^{1,2}$} \\
  \\
  $^1$ \href{https://www.columbia.edu}{Columbia University}, New York, USA \\
  $^2$ \href{https://www.recotechnologies.com}{Recognition Technologies, Inc.}, New York, USA \\
  \small{Technical Report: \href{https://www.recognitiontechnologies.com/techreport/RTI-20240822-01.pdf}{RTI-20240822-01}}\\
  \small{\href{http://dx.doi.org/10.13140/RG.2.2.32372.56963}{DOI: 10.13140/RG.2.2.32372.56963}}\\
  $^1$ {\href{mailto:bt2522@columbia.edu}{\small\texttt{bt2522@columbia.edu}}}, $^2$ {\href{mailto:beigi@recotechnologies.com}{\small\texttt{beigi@recotechnologies.com}}}
}

\date{}

\begin{document}
\maketitle

\begin{abstract}
We explore the performance of various artificial neural network architectures, including a multilayer perceptron (MLP), Kolmogorov-Arnold network (KAN), LSTM-GRU hybrid recursive neural network (RNN) models, and a time-delay neural network (TDNN) for pricing European call options. In this study, we attempt to leverage the ability of supervised learning methods, such as ANNs, KANs, and gradient-boosted decision trees, to approximate complex multivariate functions in order to calibrate option prices based on past market data. The motivation for using ANNs and KANs is the Universal Approximation Theorem and Kolmogorov-Arnold Representation Theorem, respectively. Specifically, we use S\&P 500 (SPX) and NASDAQ 100 (NDX) index options traded during 2015-2023 with times to maturity ranging from 15 days to over 4 years (OptionMetrics IvyDB US dataset). Black \& Scholes's (BS) PDE \cite{Black1973} model's performance in pricing the same options compared to real data is used as a benchmark. This model relies on strong assumptions, and it has been observed and discussed in the literature that real data does not match its predictions. Supervised learning methods are widely used as an alternative for calibrating option prices due to some of the limitations of this model. In our experiments, the BS model underperforms compared to all of the others. Also, the best MLP model outperforms the best TDNN model on all error metrics. We implement a simple self-attention mechanism to enhance the RNN models, significantly improving their performance. The best-performing model overall is the LSTM-GRU hybrid RNN model with attention. Also, the KAN model outperforms the TDNN and MLP models. We analyze the performance of all models by ticker, moneyness category, and over/under/correctly-priced percentage. Due to some of the errors being complimentary in the sense of having opposite percent over-priced and under-priced for some moneyness categories, it may be beneficial to investigate the ensembling of the best models.
\end{abstract}

\\

\myhypertarget{htag:TOC}{}
\centerline{\large {\textbf{\textit{Table of Contents}}}}
\noindent\myhyperlink{htag:Section1}{1. Problem description \& state of the art}\\
\myhyperlink{htag:Section2}{2. Derivation of BS Option Pricing PDE}\\
\myhyperlink{htag:Section3}{3. Motivation for Deep Learning \& ANNs}\\
\myhyperlink{htag:Section4}{4. Method, Data Setup, \& Expected Results}\\
\myhyperlink{htag:Section5}{5. Data Summary Statistics and Exploratory Plots}\\
\myhyperlink{htag:Section6}{6. Black-Scholes Model}\\
\myhyperlink{htag:Section7}{7. MLP Model}\\
\myhyperlink{htag:Section8}{8. XGBoost Model}\\
\myhyperlink{htag:Section9}{9. TDNN Model \& Potential Improvements}\\
\myhyperlink{htag:Section10}{10. RNN Models \& Attention Mechanism}\\
\myhyperlink{htag:Section11}{11. Kolmogorov-Arnold Representation Theorem \& KAN Model}\\
\myhyperlink{htag:Section12}{12. Evaluation \& Results}\\
\myhyperlink{htag:Section13}{13. Future Work}\\
\myhyperlink{htag:Appendix}{Appendix: Best KAN Model Parameter Count \& Equations}\\
\myhyperlink{htag:Figures}{Figures}\\
\myhyperlink{htag:References}{References}\\

\section{Problem Description \& State Of The Art}
\myhypertarget{htag:Section1}{}
Financial derivatives are contractual agreements between multiple parties, the value of which is contingent upon the performance or characteristics of a specified underlying asset (or a collection of such assets), such as stocks, market indices, interest rates, commodity prices, exchange rates, or other benchmarks \cite{OosterleeGrzelak2020}. An option contract can be defined as a financial derivative that gives the holder the right but not the obligation to buy (call) or sell (put) an underlying asset at a specified price (strike = K) on the expiration date (European style) or before it (American style). Options are traded both in the over-the-counter market and through exchanges \cite{Hull2014}. Purchasing a European call option allows one to bet on the price of the underlying asset rising since the predetermined strike price is subtracted from the terminal price of the underlying in the payoff. Hence, in the context of a European call option, the holder will rationally choose to exercise the option at the maturity time \( t = T \) if the underlying asset's price exceeds the strike price, i.e., when \( S(T) > K \). Under such circumstances, the option holder can acquire the asset by paying the strike price \( K \) to the option writer, thereby securing an asset valued at \( S(T) \). The resultant profit for the holder is \( S(T) - K \), as the asset can be immediately sold in the financial market at its current value. Conversely, if the asset's price at maturity \( S(T) \) is less than the strike price \( K \), the option holder will opt not to exercise the option, rendering it worthless \cite{OosterleeGrzelak2020}. In this scenario, the holder can purchase the asset directly in the market for a price lower than \( K \), making it irrational to utilize the contractual right provided by the option\cite{OosterleeGrzelak2020}. Notably, the holder of a European call option is not obligated to exercise the option if the underlying asset underperforms. This flexibility allows the holder to avoid unfavorable trades. However, the option seller, or writer, is contractually bound to fulfill the terms of the option should the holder choose to exercise it \cite{OosterleeGrzelak2020}. 
\\ 

A crucial aspect of what makes simple options attractive is their versatility in constructing robust trading strategies and hedging positions. By combining a collection of simple options contracts in creative ways, traders can devise sophisticated strategies that optimize returns or mitigate risks. Among the most popular strategies are the straddle, condor, butterfly, and covered call. The straddle strategy involves purchasing both a call option and a put option on the same underlying asset with the same strike price and expiration date. This strategy benefits from significant price movements in either direction, as profits from one option can offset the losses from the other, making it an effective choice in highly volatile markets \cite{Haug2007}. The condor strategy, specifically the iron condor, is a more advanced approach that involves selling an out-of-the-money put and an out-of-the-money call while simultaneously buying a further out-of-the-money put and call. This results in a strategy that profits when the underlying asset remains within a specific price range, thus limiting potential losses but also capping the maximum profit \cite{Hull2014}. The butterfly spread is another neutral options strategy that combines both calls and puts to benefit from low volatility. It involves purchasing a call (or put) at a lower strike, selling two calls (or puts) at a middle strike, and purchasing another call (or put) at a higher strike price. The butterfly spread is designed to generate a profit when the underlying asset’s price remains near the middle strike price \cite{McMillan2012}. Lastly, the covered call strategy involves holding a long position in an underlying asset while simultaneously selling a call option on the same asset. This strategy generates additional income from the premium received for selling the call, and it is particularly attractive in markets where the underlying asset is expected to experience minimal price fluctuations \cite{Natenberg1994}. 
\\

These strategies demonstrate the potential to creatively use simple options contracts to achieve diverse investment objectives, whether to speculate on price movements or to hedge position risk effectively. Since 1973, standardized options have been actively traded on regulated exchanges, while other options are privately negotiated and sold by financial institutions to their clients. A category of options known as exotic options exists, characterized by complex payoff structures, which may involve multiple underlying assets or FX rates. Aside from these complications, an exotic option may contain path dependency, where the payoff is influenced not solely by the asset price(s) at maturity \( S(T) \) or at a specific time \( S(t) \) but also by the asset's price at multiple points in time. Unlike standard options, exotic options are typically not listed on regulated exchanges but are transacted over the counter (OTC) \cite{OosterleeGrzelak2020}. This means they are customized and directly sold by banks and other financial institutions to their counterparties \cite{OosterleeGrzelak2020}.
\\

In 1973, Fischer Black and Myron Scholes co-authored a paper on arguably the most important option pricing model, which is now referred to as the BS model in short. Based on several strong assumptions, they constructed a closed-form solution to a PDE describing the evolution of European style puts and calls. The PDE used by Black \& Scholes is:

\begin{equation}
\frac{\partial f}{\partial t} + r S \frac{\partial f}{\partial S} + \frac{1}{2} \sigma^{2} S^{2} \frac{\partial^{2} f}{\partial S^{2}} - r f = 0
\end{equation}

Here, $f$ is a function of the underlying asset's price $S$, risk-free rate $r$, volatility of the underlying asset $\sigma$, and time $t$. The BS formula for the price of a vanilla European call option is:

\begin{equation}
f(t, S(t)) = S_{t} \Phi\left(d_{1}\right) - K e^{-r(T-t)} \Phi\left(d_{2}\right)
\end{equation}

\begin{equation}
d_{1} = \frac{\log \left(\frac{S_{t}}{K}\right) + (T-t)\left(r+\frac{\sigma^{2}}{2}\right)}{\sigma \sqrt{(T-t)}}
\end{equation}

and

\begin{equation}
d_{2} = d_{1} - \sigma \sqrt{(T-t)}
\end{equation}

Here, $T$, is the expiration time, $(T-t)$ is time-to-maturity, $S_{t}$ is the current price of the underlying, $K$ is the strike price, and $r, \sigma$ are as above. Also, $\Phi$ is a cumulative distribution function of a zero-mean and standard deviation 1 normal variable, meaning $\Phi(x)$ is the probability such that this variable is less than or equal to $x$. The assumptions for the B-S model (as described by J.C. Hull \cite{Hull2014}) are:

\begin{enumerate}
  \item No transaction costs or taxes

  \item No riskless arbitrage can exist

  \item No dividends or other cash flows are paid during the lifetime of the security

  \item Trading and hedging of security is continuous, with the asset's liquidity being guaranteed

  \item Risk-free rate is constant for all maturities

  \item Short selling is allowed without penalty or short rebate

  \item Stock price follows Geometric Brownian motion with volatility of the underlying asset being constant over the lifetime of the option. As outlined in \cite{OosterleeGrzelak2020}, a GBM process is described by the following Stochastic Differential Equation (SDE):

    \begin{equation}
    dS(t) = \mu S(t) \, dt + \sigma S(t) \, dB(t), \quad \text{with } S(t_0) = S_0
    \end{equation}
    
    where \( B(t) \) is a standard Brownian motion, \( \mu \) denotes the drift parameter (a constant and deterministic growth rate of the asset), and \( \sigma \) is the constant volatility parameter. Equivalently, this can be expressed in its integral form as:
    
    \begin{equation}
    S(t) = S_0 + \int_{t_0}^{t} \mu S(z) \, dz + \int_{t_0}^{t} \sigma S(z) \, dB(z)
    \end{equation}

    This model for asset price dynamics is sometimes referred to as the Samuelsen model \cite{OosterleeGrzelak2020}. 

\end{enumerate}

Although the BS model is one of the most significant pieces of mathematical finance, earning its authors the 1997 Nobel Prize in Economics, it has been widely criticized for not matching real market data. One of the earliest works that undermined the model was published by Jackwerth \& Rubinstein in 1996 \cite{JackwerthRubinstein1996}, showing that market data demonstrates skewed stock price distributions that are rarely log-normal. This goes against the BS model since it represents underlying asset prices with a log-normal distribution. A consequence of the BS model's assumption that stock price is log-normally distributed is that log returns are expected to be normally distributed. However, multiple studies have demonstrated that log return distributions are rarely normal, as real market data displays heavy-tailed distributions \cite{Cont2001}. The assumptions of constant volatility and interest rate have also been criticized as being unrealistic \cite{Gatheral2006}. Since its inception, the BS model has seen many enhancements, but despite these continued improvements, one of the BS model's persistent downfalls is its inability to capture the volatility smile. The volatility smile is a pattern where the implied volatility, inferred from market prices of options, varies with the strike price and maturity, deviating from the constant volatility assumption of the BS model \cite{Rubinstein1985, DermanKani1994}. Specifically, as discussed above, the BS model assumes that the volatility of the underlying asset remains constant over time, leading to a single implied volatility for all options on the same underlying asset. However, empirical evidence shows that implied volatilities tend to increase as options move away from the at-the-money (ATM) strike price, forming a "smile" shape when plotted against strike prices \cite{OosterleeGrzelak2020}. 
\\

Implied volatility, denoted as $\sigma_{\text{imp}}$, is the volatility value that, when inserted as a parameter into the Black-Scholes option pricing formula, reproduces the market-observed option price $f^{\text{mkt}}(K, T)$ for a given strike price $K$ and maturity $T$. Mathematically, it is defined \cite{OosterleeGrzelak2020} as:

\begin{equation}
f(t_0, S; K, T, \sigma_{\text{imp}}, r) = f^{\text{mkt}}(K, T)  
\end{equation}

where $t_0 = 0$. Here, $f(t_0, S; K, T, \sigma_{\text{imp}}, r)$ represents the theoretical option price calculated using the Black-Scholes model with the implied volatility $\sigma_{\text{imp}}$, which equates to the market price $f^{\text{mkt}}(K, T)$ observed at time $t_0 = 0$. Thus, implied volatility is extracted by using a quoted option value to recover a value for $\sigma$, which is possible since the B-S option price can be written as 1-to-1 function of volatility. Specifically, since volatility is valid in the range [ 0 , infinity), the function is restricted to this range. In fact, this function is continuous and strictly increasing in this range and has a unique solution via reasoning that relies on the absence of the arbitrage principle. Hence, by analyzing the range where this function is convex (until a point of inflection occurs), algorithms such as the Newton-Raphson method can be used to estimate values of $\sigma$ from the parameters $S, K, T, \mathrm{r}$, and observed option price $f(t_0, S; K, T, \sigma_{\text{imp}}, r)$ \cite{OosterleeGrzelak2020}. The need for such iterative methods to solve for implied volatility further underscores the limitations of the BS model.
\\

The presence of a volatility smile indicates that market participants anticipate different levels of volatility depending on the moneyness of the option. This phenomenon can be attributed to several factors, including market expectations of future volatility, supply and demand imbalances, and the heavy-tailed nature of asset return distributions, which the BS model fails to account for \cite{Ait-Sahalia2000}. In practice, the volatility smile is a significant departure from the theoretical underpinnings of the BS model and the concept of implied volatility itself, as discussed in works such as \cite{OosterleeGrzelak2020}, highlights the discrepancy between the theoretical model and market reality. The volatility smile also reflects the market's perception of risk. For instance, higher implied volatilities for out-of-the-money (OTM) options suggest that traders expect significant price movements, which are not captured by the BS model's simplistic assumptions. These expectations might be driven by anticipated events, market sentiment, or historical data showing fat tails in the distribution of returns. Overall, the existence of the volatility smile is a key indicator of the BS model's limitations and has prompted the development of more sophisticated models that better capture the complexities of real market behavior, such as stochastic volatility models \cite{Heston1993, HullWhite1987}, local volatility models \cite{Dupire1994, DermanKani1998, Coleman1999}, and jump-diffusion models \cite{Merton1976, Kou2002, KouWang2004}. Additionally, Lévy models, both finite and infinite activity \cite{Barndorff1998, Carr2002, Eberlein2001, Matache2004, Raible2000}, have been explored for their ability to better model the heavy tails and skewness observed in asset return distributions, as highlighted in \cite{OosterleeGrzelak2020}.
\\

Local volatility models, as introduced by Dupire \cite{Dupire1994} and further developed by Derman and Kani \cite{DermanKani1998}, have gained prominence as they allow for precise calibration to market-observed implied volatilities. Unlike other models that require extensive calibration of open parameters, the local volatility framework directly uses implied volatility data to match market prices of European vanilla options, thereby effectively reproducing volatility skews and smiles. This exact calibration is achieved without the need to optimize model parameters, making local volatility models particularly attractive for pricing and risk management of derivative products \cite{OosterleeGrzelak2020}. Stochastic volatility models, such as the Heston model \cite{Heston1993}, and jump-diffusion models, like those proposed by Merton \cite{Merton1976} and Kou \cite{Kou2002}, also provide more robust mechanisms for capturing market behaviors that the Black-Scholes model fails to account for, including jumps in asset prices and heavy tails in return distributions. While these models are more flexible and can be fitted to option market data, they often require intricate calibration processes to align model outputs with market prices. Moreover, the original Black-Scholes model has been further extended with enhancements such as the inclusion of dividends \cite{Merton1973} and more intricate modeling of interest rate dynamics. Notable examples include the Hull-White model, which integrates stochastic interest rates into the Black-Scholes framework to enable the pricing of interest rate-sensitive derivatives \cite{HullWhite1990}, and the Black-Derman-Toy model, which incorporates interest rate modeling to account for their effects on bond option pricing \cite{BlackDermanToy1990}. Additionally, the Heath-Jarrow-Morton (HJM) framework provides a comprehensive approach for modeling the evolution of interest rates, significantly broadening the applicability of the original Black-Scholes model in financial markets \cite{HeathJarrowMorton1992}.

\section{Derivation of Black-Scholes PDE}
\myhypertarget{htag:Section2}{}
The following derivation is based on the detailed discussion and proof by Oosterlee \& Grzelak in their widely recommended textbook \textit{Mathematical Modeling and Computation in Finance} \cite{OosterleeGrzelak2020}. An outline of the proof of the solution via the Feynman-Kac theorem is also provided in their work, but the proof is omitted in this paper. Other alternative solution methods include MC simulation, Fourier transform or characteristic function methods, discretization of the PDE via finite differences, and pricing via the martingale approach, which is also discussed by Oosterlee \& Grzelak. 

\subsection*{2.1 Itô's Lemma and Itô Process}
Itô's lemma, named after the Japanese mathematician Kiyoshi Itô, is a cornerstone in the study of stochastic processes and stochastic calculus. It provides the necessary framework for working with increments of Brownian motion \(dB(t)\) as \(dt \rightarrow 0\), functioning analogously to a Taylor expansion when handling deterministic variables and functions. One can derive solutions to SDEs and formulate pricing stochastic partial differential equations (SPDEs) for various financial derivative products through Ito's lemma. Also, it acts as a theoretical underpinning for many results in stochastic optimal control and reinforcement learning. 

First, we need to consider the following Stochastic Differential Equation (SDE) as described by Oosterlee and Grzelak \cite{OosterleeGrzelak2020}, which corresponds to the Itô process \(X(t)\):

\begin{equation}
dX(t) = \bar{\mu}(t, X(t))dt + \bar{\sigma}(t, X(t))dB(t), \quad \text{with } X(t_0) = X_0,
\end{equation}

where \(\bar{\mu}(t, x)\) and \(\bar{\sigma}(t, x)\) represent the drift and volatility functions, respectively. These functions are required to satisfy the following Lipschitz continuity conditions, which ensure that the drift and volatility functions do not exhibit rapid growth:

\begin{equation}
|\bar{\mu}(t, x) - \bar{\mu}(t, y)|^2 + |\bar{\sigma}(t, x) - \bar{\sigma}(t, y)|^2 \leq K_1 |x - y|^2
\end{equation}

\begin{equation}
|\bar{\mu}(t, x)|^2 + |\bar{\sigma}(t, x)|^2 \leq K_2(1 + |x|^2)
\end{equation}

for some constants \(K_1, K_2 \in \mathbb{R}^+\) and any \(x\) and \(y\) in \(\mathbb{R}\). When these conditions are met, it follows with probability one that a continuous and adapted solution to this SDE exists, satisfying \(\sup_{0 \leq t \leq T} \mathbb{E}[X^2(t)] < \infty\) \cite{OosterleeGrzelak2020}. Now, consider the case where a process \(X(t)\) follows the Itô dynamics outlined above, where the drift \(\bar{\mu}(t, X(t))\) and diffusion \(\bar{\sigma}(t, X(t))\) satisfy the previously mentioned Lipschitz conditions. If we define a function \(g(t, X)\) of the stochastic process \(X = X(t)\) and time \(t\), and assume that \(g(t, X)\) has continuous partial derivatives, namely \(\partial g/\partial X\), \(\partial^2 g/\partial X^2\), and \(\partial g/\partial t\), then a new stochastic variable \(Y(t) := g(t, X)\) can be shown to also follow an Itô process \cite{OosterleeGrzelak2020}. This process is governed by the same Brownian motion \(B(t)\). The result of the lemma, which makes it extremely useful for computation, is that the SPDE for \(Y(t)\) is then given by:

\begin{equation}
dY(t) = \left( \frac{\partial g}{\partial t} + \bar{\mu}(t, X)\frac{\partial g}{\partial X} + \frac{1}{2} \frac{\partial^2 g}{\partial X^2} \bar{\sigma}^2(t, X) \right) dt + \frac{\partial g}{\partial X} \bar{\sigma}(t, X)dB(t).
\end{equation}

Equivalently, Itô's lemma can also be expressed in its integral form as follows:

\begin{equation}
Y(t) = Y_0 + \int_{t_0}^t \left( \frac{\partial g}{\partial z} + \bar{\mu}(z, X)\frac{\partial g}{\partial X} + \frac{1}{2} \frac{\partial^2 g}{\partial X^2} \bar{\sigma}^2(z, X) \right) dz 
\end{equation}
\begin{equation*}
    + \int_{t_0}^t \frac{\partial g}{\partial X} \bar{\sigma}(z, X) dB(z)
\end{equation*}

This integral formulation can be particularly useful for deriving solutions to various SPDEs and evaluating complex integrals. As seen above, Itô's lemma is analogous to the Taylor expansion in standard calculus but includes an additional Itô correction term to account for the stochastic nature of the process. Specifically, the Itô correction term is the \(\frac{1}{2} \frac{\partial^2 g}{\partial X^2} \bar{\sigma}^2(t, X)\) component in the drift term, which arises due to the quadratic variation of the Brownian motion. Typically, any higher-order terms are neglected under the convention that their contribution is insignificant in the limit as \(dt \rightarrow 0\) \cite{OosterleeGrzelak2020}.

\subsection*{2.2 Black-Scholes PDE \& Solution}

Building on the assumption outlined in section 1 of this paper, Black and Scholes derived their seminal partial differential equation (PDE) for the valuation of European options \cite{Black1973}. Namely, for this derivation, we assume a constant interest rate \( r \) and volatility \( \sigma \). The market is considered liquid, meaning assets can be traded continuously and in arbitrary quantities. Additionally, short-selling is permitted without penalty, transaction costs and dividend payments are neglected, and we assume the absence of any bid-ask spread in stock/index and option prices. A fundamental task in quantitative finance is determining the fair value of a financial derivative at the time of sale, denoted as \( t = t_0 \). More generally, we seek to determine the value \( f(t, S) \) for any time \( t \geq t_0 \) and to manage the associated risk incurred when an option writer must trade the asset \( S(T) \) at maturity \( T \), given a fixed strike price \( K \). This derivation of the BS PDE is centered around the concept of a replicating portfolio, which is designed to mirror the cash flows of the financial derivative. This portfolio can be static or dynamic, with the latter requiring periodic rebalancing based on new market information. We will follow the approach in \cite{OosterleeGrzelak2020}, where a dynamic delta hedging strategy is employed, and the portfolio is continuously adjusted rather than at discrete rebalancing times. To begin, note that the underlying asset's stochastic process is assumed to be a GBM with constant volatility, which has the following aforementioned SDE under the real-world measure \( \mathbb{P} \):

$$
dS(t) = \mu S(t) dt + \sigma S(t) dB^{\mathbb{P}}(t)
$$

Given that the option price \( f(t, S) \) is a function of both time \( t \) and the stochastic process \( S(t) \), we start by applying Itô's lemma to derive its dynamics:

\begin{equation}
    df(t,S) = \frac{\partial f}{\partial t} dt + \frac{\partial f}{\partial S} dS + \frac{1}{2} \frac{\partial^2 f}{\partial S^2} (dS)^2
\end{equation}

Then, we substitute the SDE for \( dS(t) \) into the Itô expansion above, starting with the first-order terms: 

\[
df(t,S) = \frac{\partial f}{\partial t} dt + \frac{\partial f}{\partial S} (\mu S dt + \sigma S dB^{\mathbb{P}}(t)) + \frac{1}{2} \frac{\partial^2 f}{\partial S^2} (dS)^2
\]

Expanding further:

\[
df(t,S) = \frac{\partial f}{\partial t} dt + \mu S \frac{\partial f}{\partial S} dt + \sigma S \frac{\partial f}{\partial S} dB^{\mathbb{P}}(t) + \frac{1}{2} \frac{\partial^2 f}{\partial S^2} (dS)^2
\]

Next, we compute the second-order term \( (dS)^2 \), but first, we outline some conventions. 

\subsection*{2.2.1 Box Calculus Conventions for Itô Calculus}

In Itô calculus, the algebra used to handle the stochastic terms includes special rules for the manipulation of the differentials \( dt \) and \( dB_t \) and terms involving their products. In essence, we can neglect higher-order terms like \( (dt)^2 \) and \( dt dB^{\mathbb{P}}(t) \), since they become insignificant as \( dt \rightarrow 0 \). The key rules are:

\begin{equation}
    dt \cdot dt = 0, \quad dt \cdot dB_t = 0, \quad \text{and} \quad dB_t \cdot dB_t = dt
\end{equation}

For instance, when expanding the product of two differentials, say \( (i dt + j dB_t) \) and \( (\alpha dt + \beta dB_t) \), the following computation applies given these rules:

\[
(i dt + j dB_t) \cdot (\alpha dt + \beta dB_t) = 
\]

\[
i\alpha dt \cdot dt + i\beta dt \cdot dB_t + j\alpha dB_t \cdot dt + j\beta dB_t \cdot dB_t = j\beta dt
\]

Which gives the useful rule: 
\begin{equation}
    (i dt + j dB_t) \cdot (\alpha dt + \beta dB_t) = j\beta dt
\end{equation}

These conventions are crucial for the correct manipulation and simplification of stochastic differential equations in the context of Itô calculus and will be useful in simplifying the second-order term \( (dS)^2 \). Specifically, we get the following after expanding:  
\[
(dS)^2 = (\mu S dt + \sigma S dB^{\mathbb{P}}(t))^2 = \mu^2 S^2 (dt)^2 + 2 \mu \sigma S^2 dt dB^{\mathbb{P}}(t) + \sigma^2 S^2 (dB^{\mathbb{P}}(t))^2
\]

Then, the only term that remains after applying these Box calculus rules is 
\begin{equation}
    (dS)^2 = \sigma^2 S^2 (dB^{\mathbb{P}}(t))^2 = \sigma^2 S^2 dt
\end{equation}

which is simply an example of the aforementioned rule for multiplying two differentials with \(j = \beta = \sigma S\). 
Now, substituting this back into our equation for \( df(t, S) \), we get:

\[
df(t,S) = \frac{\partial f}{\partial t} dt + \mu S \frac{\partial f}{\partial S} dt + \sigma S \frac{\partial f}{\partial S} dB^{\mathbb{P}}(t) + \frac{1}{2} \frac{\partial^2 f}{\partial S^2} \sigma^2 S^2 dt.
\]

Finally, combining all terms with \( dt \), we arrive at the following expression:
\begin{equation}
    df(t,S) = \left( \frac{\partial f}{\partial t} + \mu S \frac{\partial f}{\partial S} + \frac{1}{2} \sigma^2 S^2 \frac{\partial^2 f}{\partial S^2} \right) dt + \sigma S \frac{\partial f}{\partial S} dB^{\mathbb{P}}(t)
\end{equation}

Next, we construct a replicating portfolio \( \Pi(t, S) \), consisting of a long (i.e., buy) position in one example of the option (with value \( f(t, S) \)) and a short (i.e., sell) position of size \( \Delta \) in the stock/index:

\begin{equation}
\Pi(t, S) = f(t, S) - \Delta S(t).
\end{equation}

This portfolio must hedge the risk, meaning the stochastic component governed by \( B^{\mathbb{P}}(t) \) should be eliminated. To this end, we calculate the change in the portfolio value using our existing expressions for \( df(t, S) \) and \( dS(t) \):

\begin{equation}
d\Pi = df - \Delta dS = 
\end{equation}
    
\begin{equation*}
   \left[ \frac{\partial f}{\partial t} + \mu S \frac{\partial f}{\partial S} + \frac{1}{2} \sigma^2 S^2 \frac{\partial^2 f}{\partial S^2} - \Delta \mu S \right] dt + \sigma S \left( \frac{\partial f}{\partial S} - \Delta \right) dB^{\mathbb{P}}(t)
\end{equation*}

To ensure the portfolio's value is risk-free/deterministic, we choose \( \Delta \) such that the \( dB^{\mathbb{P}} \)-terms cancel out:

\begin{equation}
\Delta = \frac{\partial f}{\partial S},
\end{equation}

Substituting this into the previous equation results in:

\begin{equation}
d\Pi = \left( \frac{\partial f}{\partial t} + \frac{1}{2} \sigma^2 S^2 \frac{\partial^2 f}{\partial S^2} \right) dt
\end{equation}

Importantly, the portfolio dynamics are now independent of the drift term \( \mu \), relying solely on the volatility \( \sigma \), which is responsible for encapsulating the uncertainty in the future behavior of stock prices \cite{OosterleeGrzelak2020}. The portfolio value, \( \Pi(t, S) \), should grow at the risk-free rate \( r \), comparable to an investment in a risk-free money savings account. This growth ensures that the return on the portfolio matches that of a risk-free investment. We can represent a bank account \( A(t) \), where the amount of money grows at the rate \( r \), as follows:

\[
A(t) = A(t_0)e^{r(t - t_0)}
\]

where \( A(t_0) \) is the initial amount in the bank account at time \( t_0 \). The differential form of the bank account growth is then given by:

\[
dA = rA \, dt
\]

For a portfolio amount \( \Pi \equiv \Pi(t, S) \), the change in the value of the portfolio, assuming it grows at the same rate \( r \), can then be expressed as:

\begin{equation}
    d\Pi = r\Pi \, dt
\end{equation}

Here, \( r \) represents the constant interest rate corresponding to a risk-free savings account, and the equation above highlights that the portfolio's value increases at the risk-free rate, consistent with a no-arbitrage condition.
\\

Combining equations 18, 20, and 22, we get:

\begin{equation}
d\Pi = r \left( f - S \frac{\partial f}{\partial S} \right) dt
\end{equation}

Finally, equating equations 21 and 23, and simplifying, we arrive at the Black-Scholes PDE (equation 1) for the option value \( f(t, S) \):

\[
\frac{\partial f}{\partial t} + r S \frac{\partial f}{\partial S} + \frac{1}{2} \sigma^2 S^2 \frac{\partial^2 f}{\partial S^2} - r f = 0 
\]

This is a parabolic PDE, with the "+"-sign in front of the diffusion term \( \frac{1}{2} \sigma^2 S^2 \frac{\partial^2 f}{\partial S^2} \), indicating the problem is well-posed when accompanied by a final condition. As explained in \cite{OosterleeGrzelak2020}, the final condition is usually provided by the payoff function \( H(T, S) \), where \( f(T, S) = H(T, S) \) since the discount factor reduces to 1 at time t = T. For a European vanilla call option, the payoff has the following form:

\begin{equation}
    H(T, S) := \max(S(T) - K, 0),
\end{equation}

and the value of the option is the continuously discounted expectation of the payoff under the risk-neutral measure. This is outlined concisely in the statement of the Feynman-Kac theorem tailored to this problem presented by Oosterlee \& Grzelak \cite{OosterleeGrzelak2020}:

\subsection*{2.2.2 Feynman-Kac Theorem for BS PDE}
Given the money-savings account, modeled by \( dA(t) = r A(t) dt \), with constant interest rate \( r \), let \( f(t, S) \) be a sufficiently differentiable function of time \( t \) and stock price \( S = S(t) \). Suppose that \( f(t, S) \) satisfies the following partial differential equation, with general drift term, \( \bar{\mu}(t, S) \), and volatility term, \( \bar{\sigma}(t, S) \):

\begin{equation}
    \frac{\partial f}{\partial t} + \bar{\mu}(t, S) \frac{\partial f}{\partial S} + \frac{1}{2} \bar{\sigma}^2(t, S) \frac{\partial^2 f}{\partial S^2} - r f = 0,
\end{equation}

with a final condition given by \( f(T, S) = H(T, S) \). The solution \( f(t, S) \) at any time \( t < T \) is then given by:

\begin{equation}
    f(t, S) = e^{-r(T-t)} \mathbb{E}^\mathbb{Q} \left[ H(T, S) \mid \mathcal{F}(t) \right] =: A(t) \mathbb{E}^\mathbb{Q} \left[ \frac{H(T, S)}{A(T)} \mid \mathcal{F}(t) \right],
\end{equation}

where the expectation is taken under the measure \( \mathbb{Q} \). 
\\

According to the Feynman-Kac theorem, the task of solving the Black-Scholes partial differential equation (PDE), which emerges by selecting the drift term $\bar{\mu}(t, S) = rS$ and the diffusion term $\bar{\sigma}(t, S) = \sigma S$, can be reformulated as the computation of the expected value of a discounted payoff function under the $\mathbb{Q}$-measure, as explained in \cite{OosterleeGrzelak2020}. 

\subsection*{2.2.3 Derivation of BS Formula}
Hence, this final condition is necessary to solve the PDE, and the closed-form solution in equations 2-4 of this paper can be recovered as follows:

\begin{equation}
f(t,S) = e^{-r(T-t)} \mathbb{E}^{\mathbb{Q}} \left[ \max(S(T) - K, 0) \mid \mathcal{F}(t) \right]
\end{equation}

\noindent This can be decomposed into two parts using indicator functions:

\begin{equation}
f(t,S) = 
\end{equation}
\begin{equation*}
    e^{-r(T-t)} \mathbb{E}^{\mathbb{Q}} \left[ S(T) \mathbf{1}_{\{S(T) > K\}} \mid \mathcal{F}(t) \right] - e^{-r(T-t)} \mathbb{E}^{\mathbb{Q}} \left[ K \mathbf{1}_{\{S(T) > K\}} \mid \mathcal{F}(t) \right]
\end{equation*}

\noindent Before proceeding, we review some valuable properties of Lognormal random variables as discussed in \cite{Rouah2013}. 

\subsection*{The Lognormal PDF and CDF}
Firstly, we use the fact that if a random variable $Y \in \mathbb{R}$ follows the normal distribution with mean $\mu$ and variance $\sigma^2$, then $X = e^Y$ follows the lognormal distribution with mean
\[
E[X] = e^{\mu + \frac{1}{2}\sigma^2}
\]
and variance
\[
Var[X] = \left( e^{\sigma^2} - 1 \right) e^{2\mu + \sigma^2}.
\]
The pdf for $X$ is
\begin{equation}
    dF_X(x) = \frac{1}{\sigma x \sqrt{2\pi}} \exp\left(-\frac{1}{2} \left(\frac{\ln x - \mu}{\sigma}\right)^2\right)
\end{equation}

\noindent and the cdf is

\begin{equation}
    F_X(x) = \Phi \left( \frac{\ln x - \mu}{\sigma} \right)
\end{equation}

\noindent where $\Phi(y) = \frac{1}{\sqrt{2\pi}} \int_{-\infty}^{y} e^{-\frac{1}{2}t^2} dt$ is the standard normal cdf.

\subsection*{The Lognormal Conditional Expected Value}
As in \cite{Rouah2013}, we let the expected value of $X$ conditional on $X > x$ be $L_X(K) = E[X|X > x]$. For the lognormal distribution, using Equation 29, this becomes:
\begin{equation}
    L_X(K) = \int_{K}^{\infty} \frac{1}{\sigma \sqrt{2\pi}} e^{-\frac{1}{2} \left(\frac{\ln x - \mu}{\sigma}\right)^2} dx
\end{equation}

\noindent Then, with the change of variable $y = \ln x$, we get $x = e^y$, $dx = e^y dy$, and the Jacobian is $e^y$. Hence we have

\begin{equation}
    L_X(K) = \int_{\ln K}^{\infty} \frac{e^y}{\sigma \sqrt{2\pi}} e^{-\frac{1}{2} \left(\frac{y - \mu}{\sigma}\right)^2} dy
\end{equation}

\noindent Combining terms and completing the square, the exponent is
\[
-\frac{1}{2\sigma^2} \left( y^2 - 2y\mu + \mu^2 - 2\sigma^2 y \right) = -\frac{1}{2\sigma^2} \left( y - (\mu + \sigma^2) \right)^2 + \mu + \frac{1}{2}\sigma^2.
\]
Equation 32 becomes
\begin{equation}
    L_X(K) = \exp\left(\mu + \frac{1}{2}\sigma^2\right) \frac{1}{\sigma} \int_{\ln K}^{\infty} \frac{1}{\sqrt{2\pi}} \exp\left(-\frac{1}{2} \left( \frac{y - (\mu + \sigma^2)}{\sigma} \right)^2\right) dy
\end{equation}

Consider the random variable $X$ with pdf $f_X(x)$ and cdf $F_X(x)$, and the scale-location transformation $Y = \sigma X + \mu$. It is easy to show that the Jacobian is $\frac{1}{\sigma}$, that the pdf for $Y$ is $f_Y(y) = \frac{1}{\sigma}f_X\left(\frac{y-\mu}{\sigma}\right)$ and that the cdf is $F_Y(y) = F_X\left(\frac{y-\mu}{\sigma}\right)$. Hence, the integral in Equation 33 involves the scale-location transformation of the standard normal cdf. Using the fact that $\Phi(-x) = 1 - \Phi(x)$ this implies that
\begin{equation}
    L_X(K) = \exp\left(\mu + \frac{\sigma^2}{2}\right) \Phi\left(-\frac{\ln K + \mu + \sigma^2}{\sigma}\right)
\end{equation}

\noindent Now, we return to equation 28 and compute the two terms of conditional expectations to recover equations 2-4. We use the same approach as in \cite{Rouah2013}. 
\[
f(t,S) = e^{-r(T-t)} \mathbb{E}^{\mathbb{Q}} \left[ S(T) \mathbf{1}_{\{S(T) > K\}} \mid \mathcal{F}(t) \right] -  
\]
\[
e^{-r(T-t)} \mathbb{E}^{\mathbb{Q}} \left[ K \mathbf{1}_{\{S(T) > K\}} \mid \mathcal{F}(t) \right] =
\]

\begin{equation}
    e^{-r\tau} \int_{K}^{\infty} S_T dF(S_T) - e^{-r\tau} K \int_{K}^{\infty} dF(S_T)
\end{equation}

\noindent with
\[
    \tau = (T - t) 
\]

\noindent To evaluate the two integrals, we make use of the result derived above that under $\mathbb{Q}$ and at time $t$ the terminal stock price $S_T$ follows the lognormal distribution with mean $\ln S_t + \left(r - \frac{\sigma^2}{2}\right)\tau$ and variance $\sigma^2\tau$, where $\tau$ is the time to maturity.

\begin{equation}
    S(T) \sim \text{Lognormal}\left( \log S_t + \left( r - \frac{\sigma^2}{2} \right)(T-t), \sigma^2 (T-t) \right)
\end{equation}

\noindent The first integral uses the conditional expectation of $S_T$ | $S_T > K$

\begin{equation}
    \int_{K}^{\infty} S_T dF(S_T) = \mathbb{E}^Q [S_T \mid S_T > K] = L_{S_T}(K)
\end{equation}

\noindent This conditional expectation is, from Equation 34
\begin{equation}
    L_{S_T}(K) = \exp\left(\ln S_t + \left(r - \frac{\sigma^2}{2}\right)\tau + \frac{\sigma^2\tau}{2}\right) \times 
\end{equation}
\begin{equation*}
    \Phi\left(-\frac{\ln K + \ln S_t + \left(r - \frac{\sigma^2}{2}\right)\tau + \sigma^2\tau}{\sigma \sqrt{\tau}}\right) 
\end{equation*}

\noindent which simplifies to 
\begin{equation}
    L_{S_T}(K) = S_t e^{r\tau} \Phi(d_1)
\end{equation}

\noindent so the first integral is
\[
 e^{-r\tau} L_{S_T}(K) = e^{-r\tau} S_t e^{r\tau} \Phi(d_1) = S_t \Phi(d_1)
\]
Using Equation 30, the second integral in Equation 35 can be written
\begin{equation*}
    e^{-r\tau} K \int_{K}^{\infty} dF(S_T) = e^{-r\tau} K [1 - F(K)] = 
\end{equation*}

\begin{equation}
    e^{-r\tau} K \left[1 - \Phi\left(\frac{\ln K - \ln S_t - \left(r - \frac{\sigma^2}{2}\right)\tau}{\sigma \sqrt{\tau}}\right)\right]
\end{equation}

\begin{equation}
    = e^{-r\tau} K [1 - \Phi(-d_2)] = e^{-r\tau} K \Phi(d_2)
\end{equation}

\noindent Thus, combining the two terms, the equation simplifies to:

\[
f(t,S) = S_t \Phi(d_1) - e^{-r(T-t)} K \Phi(d_2)
\]

\noindent where \( d_1 \) and \( d_2 \) are defined as:

\[
d_1 = \frac{\log\left(\frac{S_t}{K}\right) + (T-t)\left(r + \frac{\sigma^2}{2}\right)}{\sigma \sqrt{T-t}}
\]

\[
d_2 = d_1 - \sigma \sqrt{T-t}
\]

\noindent Therefore, we recover equations 2-4. These steps outline the derivation of the Black-Scholes formula, where the terms correspond to the expected payoff of the option under the risk-neutral measure \( \mathbb{Q} \), discounted to the present time. 

\section{Motivation for Deep Learning \& ANNs}
\myhypertarget{htag:Section3}{}
The application of artificial neural networks (ANNs) to option pricing gained traction in the 1990s. One of the pioneering studies, conducted by Malliaris and Salchenberger \cite{Malliaris1993}, employed a multilayer perceptron (MLP) network to estimate the prices of S\&P 100 call options. Their findings demonstrated that the neural network frequently outperformed the BS model in terms of prediction accuracy, as measured by mean squared error (MSE). Since then, a diverse array of ANN architectures and deep learning methodologies have been explored for option pricing, facilitated by the availability of vast amounts of market data for model training. The Universal Approximation Theorem \cite{Hornik1989, Csaji2001, Leshno1993} asserts that feed-forward neural networks possess the capacity to approximate a vast class of functions through the learning of suitable weights. This foundational principle underpins the motivation behind this study and similar efforts aimed at deriving the functional form of option pricing via neural network training. A particularly influential contribution in this domain was made by Hutchinson et al. \cite{Hutchinson1994}, whose 1994 paper presented a nonparametric approach to pricing and hedging derivative securities using learning networks. Their study focused on S\&P 500 futures options and similarly reported superior performance compared to the BS model. An extensive review of such applications of neural networks to option pricing and hedging is provided by Ruf and Wang \cite{RufWang2020}, offering a comprehensive overview of developments up to 2020. More recently, Ferraz \cite{Ferraz2022} employed the XGBoost algorithm, incorporating the parameters of the BS model as inputs, to predict option prices. Furthermore, with advancements in the field, even Physics-Informed Neural Networks (PINNs), originally devised to solve partial differential equations (PDEs) in physics, have been adapted for derivative pricing under the right conditions \cite{Dhiman2023}.
\\

Moreover, research has also demonstrated the ability of neural networks to capture the volatility smile. For instance, studies by Dugas et al. \cite{Dugas2009} and Liu et al. \cite{Liu2019} have employed deep learning architectures that effectively model the volatility smile by capturing the complex non-linear relationships between option prices and their determinants. Dugas et al. showed that neural networks could better replicate market prices than the BS model, particularly in scenarios where the implied volatility surface exhibits strong curvature. Similarly, Liu et al. demonstrated that their model, incorporating volatility surface features, outperformed traditional models by capturing the nuanced effects of asymmetry and tail risk in return distributions. The inputs for the ANN or deep learning model can vary but tend to be \cite{Pohjonen2022} some subset of the following: past underlying prices $S$, strike price $K$, time-to-maturity $(T-t)$, risk-free rate r, and volatility estimates. Inputs such as volume, skewness, kurtosis, and others have also been used before. There are multiple alternatives for producing the volatility estimates, with some noteworthy ones being GARCH (generalized autoregressive conditional heteroskedasticity) model volatility forecasts as done in \cite{Pohjonen2022} and historical realized volatilities for different time scales. Hull describes the calculation of realized volatility \cite{Hull2014} from historic returns over a time period $\tau$ as:

\begin{equation}
    \widehat{\sigma}_{\tau}=\frac{1}{\sqrt{\tau}}\left(\sqrt{\left.\frac{1}{n-1} \sum_{i=1}^{n}\left(u_{i}-\bar{u}\right)^{2}\right)}\right.
\end{equation}

Here, $u_{i}=\ln \left(\frac{s_{i}}{s_{i-1}}\right), \mathrm{n}+1$ is the number of observations, $S_{i}$ is the stock price at the end of $\mathrm{i}^{\text {th }}$ interval, and $\tau$ is the length of the time interval in days. For example, for 1 year, the standard assumption is 252 trading days, meaning $\tau=252$. As explained by T. Pohjonen: 'According to Merton (1973), the return of the underlying asset is independent of the level of the price of the underlying asset $S$ such that it is also independent of the pricing function $f(\cdot)$ of an option price $C$. Therefore, ANNs are trained to estimate the price \( C \) divided by the strike \( K \). This leads us to the functional form:

\begin{equation}
    \frac{C}{K} = f\left(\frac{S}{K}, \frac{K}{K}, T-t, r, \text{volatility estimates...}\right)
\end{equation}

where some number of volatility estimates can be chosen \cite{Pohjonen2022}. Other alternative approaches that are widely used in the literature include applying some form of smoothing (e.g., exponential) to the historic volatility estimates, interpolating interest rate estimates from yield curves, and using implied volatility estimates or modeling volatility as stochastic. 

\section{Method, Data Setup, \& Expected Results}
\myhypertarget{htag:Section4}{}
In this study, the aforementioned functional form is used with six volatility estimates of $20,30,40,50,65$, and 90 days all annualized assuming 252 trading days in a year. Only historic volatility estimates are used without implied or GARCH-fitted volatility. The focus is restricted to cash-settled European-style call options for non-dividend paying indices of S\&P 500 (SPX), Dow-Jones Index (DJX), and NASDAQ 100 (NDX). This choice of options should ensure adherence to BS's intended use case of European exercise style and no dividends. A majority of studies tackling the problem of option pricing using ANNs and other deep learning methods focus on S\&P 500 options. Hence, this selection allows for a good comparison to previous work as well as expanding the range of indices addressed through the inclusion of NDX and DJX. As described in \cite{Pohjonen2022}, options with time-to-maturity $<15$ days, $\mathrm{S} / \mathrm{K}<0.8$ or $\mathrm{S} / \mathrm{K}>1.2$, and $C<S-K e^{-r * r}$ are excluded from consideration. Such options are low-priced, deep-in- or deep-out-of-money, or not consistent with the no-arbitrage assumption (respectively), meaning they could lead to large deviations between theoretical and observed prices, or they have little informational content as they are rarely traded. Options are called at-the-money (ATM) when the strike price is approximately equal to the price of the underlying (the ratio $\mathrm{S} / \mathrm{K}$ is close to 1 ) and in-the/out-of-the-money when $\mathrm{K}<\mathrm{S}$ and $\mathrm{K}>\mathrm{S}$ respectively. The $\mathrm{B}\mathrm{S}$ model is used as a benchmark with MAE (mean absolute error), MSE (mean squared error), and RMSE (root mean squared error) metrics calculated to check its performance on real market data. Then, different multilayer perceptron architectures are tested with automatic hyperparameter tuning by trying different architectures and combinations of parameters. After the MLP NNs, XGBoost is tested with automatic parameter tuning. Finally, after good versions of the MLP and XGBoost models are found, we move on to the KAN, TDNN and RNN models.

\subsection*{4.1 Data Setup}
The data for this study comes from the Wharton Research Data Services as a subset of the (optionm\_all) IvyDB US by OptionMetrics data set. The subset collected is called df for simplicity. Specifically, it includes only European calls for the three indices selected for the date range: 2015/05/13 $2023 / 02 / 28$. The data includes columns for date and exdate, which can be used to calculate the time-to-expiration. The C column is calculated as the average of the best bid and best offer prices for the call option as done in \cite{Ferraz2022}. The documentation of the data indicates that the 'strike price' column is actually $1000 * \mathrm{~K}$, so the K column can be created by dividing this 'strike price' by 1000. The 'target\_C/K' column is then found by dividing the calculated C by the calculated K. Data for the underlying indices' closing prices (daily) is collected from Yahoo Finance for the same date range. The column ' $S$ ' is created by filling in the rows of df with the appropriate closing price for the underlying security based on the date and ticker columns. After this, the ' S ' column is adjusted to be ' $\mathrm{S} / \mathrm{K}$ ' by dividing the old values by K. The columns 'sigma\_w' for $\mathrm{w}=\{20,30,40,50,65,90\}$ are added by the aforementioned method to include rolling window historic volatility estimates. The 13-week US Treasury Bill rates (collected from Yahoo Finance) are used for r without smoothing. Before filtering, this data set includes 23,187,972 rows/observations. After filtering (df\_filtered) and removing rows that caused missing values after volatility estimate calculations, $3,794,217$ rows remain with only two of the indices (SPX and NPX). The remaining data consists of $56.1 \%$ and $43.9 \%$ observations of SPX and NDX, respectively. This includes all dates available and all calls meeting the filtering criteria. Our filtered data contains 148, 897 different options contracts (unique option\_id). The data set is split with the first $70 \%$ samples being training and the remaining $30 \%$ being split evenly into validation and test sets. Moreover, the ranges $[0.95,1.05],[0.8,0.95)$, and $(1.05,1.2]$ are used for ATM, OTM, and ITM moneyness ratio classification respectively. Also, predictions within a $5 \%$ margin around the actual $\mathrm{C} / \mathrm{K}$ ratio are classified as being correctly priced, with predictions below and above this range being labeled as under and overpriced, respectively. These ranges for ATMs and correctly priced can potentially be narrowed to improve evaluation.

\subsection*{4.2 Expected Results}
We anticipate slight differences in performance across the two remaining tickers, with the best results expected for SPX, as it accounts for the largest proportion of the training data after filtering. Additionally, we expect the models to perform better on SPX than on NDX, given SPX's more even distribution of proportions by moneyness level, as well as its lower range and standard deviation of call prices. Moreover, based on the literature, we expect superior performance on in-the-money and out-of-the-money calls, as these options are traded more frequently in practice compared to at-the-money options. 

\section{Data Summary Statistics \& Exploratory Plots}
\myhypertarget{htag:Section5}{}
As seen in Figure 1, the data does not reflect the BS model's assumption that the risk-free rate is constant throughout the lifetime of an option. The dashed red vertical lines (30 days apart) on this plot (showing interest rate over time) clearly indicate that, even within 1-month periods, interest rate fluctuates significantly. Moreover, as shown in Figure 2, for both tickers, historical volatility estimates fluctuate a lot, with these changes being less erratic for longer window sizes. Hence, the BS model's assumption that volatility is constant throughout the lifetime of an option is also not met. These two observations are why methods such as smoothing are used in the literature. As shown in Figure 3, log returns distributions for prices of both indices are not exactly normal. Specifically, for both tickers, heavier tails and a sharper peak can be seen, which resembles Laplace distributions. Hence, the data does not demonstrate all three of these assumptions of the BS model. As shown in Figure 4, the data has a larger proportion of OTM than ITM calls for both tickers and significantly fewer ATM tickers for NDX than SPX. Indeed, as mentioned before, options for the NDX ticker have a much more uneven distribution of moneyness proportions. Out of the two tickers, SPX has the largest proportion of ATM calls, and NDX has the largest proportion of OTM calls. In addition, Figure 4 also shows that NDX calls are more expensive than SPX ones across all moneyness levels, and NDX has a wider range and larger standard deviation of call prices. Aside from the proportions of the data made up by each ticker, these observations about moneyness also support the expectation that all models should do better on the SPX ticker. The shortest time to expiration in the data is 15 days, with the longest being approximately 1800 days. However, as shown in Figure 5, most of the calls in the data have $<60$ time to expiration (days). For both tickers, ITM calls have a larger price on average, with the second largest being ATM calls and OTM calls being the cheapest. Also, Figure 5 shows that for calls with a longer time to expiration, the average price for each moneyness category is higher. As shown in the last figure, the range of K is approximately 3.5 times larger for NDX than SPX call options (top subplot). This happens since the NDX index price range for 2015-2023 is approximately 4 times wider than the SPX index price range (as shown in the bottom subplot) and K is subtracted from the underlying asset's price in a call option payoff. Furthermore, our dataset has contracts on both indices with an expiry (T) of up to 4 years but only for the SPX index with expiry of 4.5 - 7+ years. As seen in the plot there are gaps for this section of SPX call option contracts with T > 4.5 years, which reflects that such contracts are traded less regularly.  
\\

\section{Black-Scholes Model}
\myhypertarget{htag:Section6}{}
The results of fitting the BS model are good - as expected. Still, it is noteworthy that many studies in the literature achieved better results with this model by putting more work into the quality of its inputs. In this study, relatively simple estimates are used for volatility and risk-free rate. The BS model is restricted to predicting the test set for fair comparison with the other models. Also, since the output of the B-S model is call option price but the output of the other models is the ratio of this price over strike price, the BS model's predictions were divided by K for comparison. As shown in Figure 6, for the volatility estimates, wider window sizes yield more accurate pricing with the BS model. Across all error metrics, the best-performing version of the BS model is the one with 90-day historical volatility. This raises the question of whether testing even wider window sizes would improve the BS model's predictions. The best version of the BS model does significantly better on the SPX ticker than on NDX, as expected. Figure 7 shows a plot of the 90-day window size volatility BS model's predictions vs actual values of the $\mathrm{C} / \mathrm{K}$ ratio. Overall, this model yields predictions that are $59.29 \%$ overpriced, $23.72 \%$ under-priced, and $16.98 \%$ correctly priced (within $5 \%$ margin of actual C/K value), meaning that its main weakness is overpricing calls. Although, as mentioned in section 3, it may be beneficial to test narrower ranges for the 'correctly priced' margin.
\\

\section{MLP Model}
\myhypertarget{htag:Section7}{}
For the MLP models, as well as the XGBoost, TDNN, and RNN models, the same set of features is used as inputs. Namely, past underlying prices divided by strike price $S / K$, strike price $K$, time-to-maturity $(T-t)$, risk-free rate r, and all six historic volatility estimates. The target for all of these models is the ratio of call option price over strike price $\mathrm{C} / \mathrm{K}$. All the MLP architectures tested have 10 neurons in the input layer (to match the number of features) and a single neuron in the output layer to produce the predictions. The Adam optimizer with MSE loss function is used for training. The MSE loss function has the following form: 
\begin{equation}
    \frac{1}{N} \sum_{i=1}^{N}\left(p_{i} - a_{i}\right)^{2}
\end{equation}
where $\mathrm{p}_{\mathrm{i}}$ and $\mathrm{a}_{\mathrm{i}}$ are the $\mathrm{i}^{\text {th }}$ predicted and actual $\mathrm{C} / \mathrm{K}$ values respectively. Additionally, an early stopping criterion based on validation loss not decreasing for 10 epochs (restoring best weights from before stopping) is used to prevent overfitting. All variants of the model were trained for 100 epochs, and all combinations of the following architectures and parameters (respectively) were tested: $\{$ neurons $=[32,64,128,256]$, layers $=[2,3,4]\}$, \{activations $=$ ['relu', 'tanh', 'sigmoid'], learning rates $=$ $[0.00008,0.0001,0.00015,0.0005]\}$.
\\

The best-performing MLP model has 3 layers (excluding input and output layers) of 64 neurons, each with tanh activations throughout and a learning rate of 0.00008. It is trained for 100 epochs with the same early stopping criterion as mentioned before, after which it is fit on the test set to yield predictions. Figure 9 shows the best MLP model's training and validation loss over epochs. As evident in the plot, the validation loss fluctuates a lot towards the end of training, so it may be beneficial to train for longer with a larger patience for early stopping. Despite this, as seen in Figure 10, the best MLP model's predictions for $\mathrm{C} / \mathrm{K}$ fit the actual values much more tightly, and it does not have such a weakness with overpricing as the B-S model, with much more even proportions of prediction below and above the actual values. But, for some reason, it seems to struggle with calls that have larger values of this ratio. Specifically, it tends to under-price these calls more than ones with a lower value of this ratio. In fact, the best MLP model yields $27.59 \%$ overpriced, $44.21 \%$ under-priced, and $28.20 \%$ correctly priced calls. As expected, the MLP model also does better on the SPX ticker than on NDX. As shown in Figure 12, which shows the predictions and actual values across time, the B-S model struggles more toward the end of the test set. As shown in Figure 13, the best MLP model also has slightly worse performance towards the end of the test set, but this difference is much less noticeable than for the B-S model. In fact, as shown in Figure 14, the best MLP model did better than the best B-S model across all error metrics.
\\

The simplest potential improvement to the MLP model is adding dropout (or Gaussian dropout) or 11/12 regularization, which has not been implemented yet. Batch/layer normalization or various weight initializations, such as Xavier or He initialization, could also improve the training. The swish, GELU (both shown in figure 15), SwiGLU activations, or some adaptive activation functions could be tested as well. Also, it could be beneficial to test MLP architectures with different activations in different layers, as the same activation was used in all layers when testing different activations for this project. However, the potential improvement that could yield the biggest difference in performance would be to add GRU (gated recurrent unit) or LSTM layers to handle the temporal dependence in the data, which is discussed in section 10. Finally, as done in \cite{Pohjonen2022}, the final output layer could have an exponential function to ensure the predicted values are positive since $\mathrm{C} / \mathrm{K}$ values cannot be below 0.
\\

\section{XGBoost Model}
\myhypertarget{htag:Section8}{}
XGBoost stands for extreme gradient boosting, and it combines lots of weaker learners (decision trees) into a strong learner. When used for prediction/regression, decision trees work by systematically breaking down the dataset into smaller and more specific subsets, which starts at the root of the tree. The root node encompasses the entire dataset, representing the initial, undivided data. The algorithm then engages in feature selection for splitting, where it identifies a particular feature and a specific point on that feature to bifurcate the data into two segments, as shown in Figure 16. This decision is guided by the goal of minimizing mean squared error in the case of a continuous target variable. Following this, branches are created. The dataset is divided into two groups, each determined by the selected feature and split point, leading to the formation of two new nodes. This step marks the initial separation of the data based on distinct feature values and the determined cut-off value. The process of recursive splitting then continues, and this methodology of splitting is applied repeatedly to each resulting branch. The algorithm selects the features and split points that contribute to the largest reduction in the variance of the target variable within these newly formed subsets. An important aspect of this process is the stopping criteria. The division of data continues until the maximum tree depth is reached, but other alternatives for the stopping criteria exist. Prediction within a decision tree is executed at the leaf nodes. Each leaf node is responsible for making a prediction, typically calculated as the mean of the feature values of the observations grouped under that node. When a new data point is introduced for prediction, it 'moves' down the tree, following the paths determined by its feature values, until it reaches a leaf node. The mean target value of this final node is then used as the predicted value.
\\

As mentioned, XGBoost grows each tree up to the specified maximum depth, and then the trees are pruned back to their optimal size. XGBoost uses bagging (bootstrap aggregating) and boosting to reduce errors caused by variance and bias, respectively. Bagging is training each tree on different random subsets of the data and then using a weighted sum of the individual tree predictions for the overall prediction, as shown in Figure 17. Boosting is adding each tree sequentially to correct errors made by previously added trees. Gradient boosting minimizes the loss function by adding the trees using a gradient-descent-like procedure. The XGBoost implementation used has many parameters that were not tested, such as subsample and colsample\_by\_tree, both of which are built in ways of reducing overfitting. As described in the documentation, subsample specifies the fraction of the training data to be randomly sampled for each tree, and colsample\_bytree controls the fraction of features to be randomly sampled for each tree. Subsampling occurs once in each boosting iteration for subsample and once for every tree constructed for colsample\_bytree. Both of these parameters have default values of 1, which are used in this project. This means that each tree is trained on all of the data and all of the features. Different values of the max\_depth, n\_estimators, alpha, learning rate, and lambda are tested. The documentation explains the max\_depth parameter as the stopping criterion for the maximum depth of each tree that can be reached, n\_estimators as the number of trees used, and alpha/lambda as the coefficients for $11 / 12$ regularization on the weights. It is important to note that the same random\_state is used in the sampling for all XGBoost models tested to ensure fair comparison, which would be especially crucial if the subsample parameter value were varied from 1. All combinations of the following architectures and parameters (respectively) were tested: $\{$ max\_depth $=[5,10,20,35,55$, 75], n\_estimators $=[50,100,500,1000,3000]\},\{$ alpha $=$ $[0,1 e-5,1 e-4,1 e-3,1 e-2]$, lambda $=[0,1 e-5,1 e-4,1 e-3]$, learning rates $=[0.1,0.2,0.3,0.4]\}$. The learning rates range was tweaked with preliminary training stages before testing all combinations of the aforementioned options.
\\

The best-performing XGBoost model has 50 trees and a max\_depth value of 35 with $0.1,0.01$ and 0.00001 for the learning rate, alpha, and lambda values, respectively. This model was trained with early stopping based on the validation set error with a patience of 10 boosting rounds. Figure 18 shows the training and validation errors over boosting rounds for this model. As seen, the training and validation error curves are nearly identical and decrease sharply at the beginning, then flatten out as the number of boosting rounds increases. This indicates that we avoid overfitting as, typically, overfitting would show a divergence between the two curves, where the training error continues to decrease while the validation error starts increasing or being unstable after a certain number of rounds. Indeed, as shown in Figure 19, the best XGBoost model did surprisingly well with its predictions for $\mathrm{C} / \mathrm{K}$, fitting the actual values much more tightly than the best BS, MLP, and TDNN models. Overall, the best XGBoost model yielded $19.03 \%$ overpriced, $38.97 \%$ under-priced, and $42.00 \%$ correctly priced calls. Indeed, as shown in Figure 21, this model also handled the later points of the test set better than the previous models. Figure 22 is a table of the error metrics by model for XGBoost, MLP, and B-S. Also, Figure 27 shows the error metrics by model for all of the models. It is evident that XGBoost outperforms the best MLP, TDNN, KAN, and B-S models across all metrics. 
\\

\section{TDNN Model \& Potential Improvements}
\myhypertarget{htag:Section9}{}
TDNN stands for Time-Delay Neural Network, and this model is especially well suited to handle temporal dependence in the data due to its architecture and handling of the features. The architecture of TDNN is characterized by its hierarchical structure, which employs varying temporal convolutions. This structure allows each layer in the network to establish connections that span across outputs from the preceding layer, which improves the network's ability to capture temporal dynamics. As the network processes data deeper into its layers, each unit within the network effectively broadens its 'view' or 'scope' of the input sequences. This expanding receptive field allows the network to incorporate a wider context at each subsequent layer, gaining an enhanced understanding of temporal patterns within the data. Specifically, in the TDNN, different layers handle different time steps, such as $\mathrm{t}-3, \mathrm{t}-2, \mathrm{t}, \mathrm{t}+2$, $t+3$. This approach is tailored to address varying temporal dependencies, allowing the network to focus on different segments of the time series data in a more targeted manner. Another critical feature of TDNN is the inclusion of statistical pooling (based on mean and standard deviation) across segments of the input sequence. This pooling mechanism serves to summarize the extracted features. By doing so, it ensures that the relevant characteristics of the sequences are efficiently captured and distributed throughout the network. The overall architecture of the TDNN can be summarized as having three main sections: (1) the frame-level layers, which consist of temporal convolutions; (2) the statistical pooling layer, which condenses the outputs from the frame-level layers; and (3) the segment level layers, which further refine these features and include an embedding layer followed by the prediction layer.
\\

The actual TDNN model is complex and takes excessive tuning. Due to the time constraints of this project, a simpler TDNN-inspired neural network model is used, with the main differences being the absence of statistical pooling and a vastly simplified hierarchal structure of temporal convolutions. The current model structure is a series of sequential 1D convolution layers (Conv1D from Keras) with the same context window and kernel sizes throughout. All of these Conv1D layers also have padding = 'same' to ensure the inputs to layers following the first one are of the appropriate dimension. Also, all of these layers have the same activations throughout, and each of them is followed by a dropout regularization layer with the same proportion throughout (of the preceding layer's outputs, which are randomly set to 0 ). After the temporal convolution and dropout layers, the output is flattened and fed into a dense layer with a single neuron for prediction. Also, the same reshaping is applied to all the feature sequences. This reshaping must be applied to the data before it can be fed into the TDNN-like model. Specifically, the following function was used to perform this data manipulation:

\begin{verbatim}
    def reshape_data(X, time_steps):
    
        reshaped = []
    
        for i in range(len(X) - time_steps + 1):
            reshaped.append(X[i: i + time_steps])
        
        return np.array(reshaped)
\end{verbatim}

This function is designed to transform a two-dimensional array X, representing feature time sequences, into a three-dimensional array, with each element representing a time-windowed sequence, which can then be fed into the TDNN-like model. The parameter time\_steps is an integer that determines the length of each temporal slice in the reshaped data. The function works by iterating through the array X and, for each iteration indexed by $i$, it extracts a slice of the data starting from the current index i and extending time\_steps rows 'forward'. This slice includes the current time step as well as the past \#(time\_steps - 1) time steps. As a result, each slice forms a sequence that captures a specific window of time in the data. As mentioned, the output of this function is a three-dimensional array. This array is essentially a compilation of the extracted slices, where each slice represents a sequence of data corresponding to a particular feature over the specified time window.
\\

As with the MLP model, the Adam optimizer and MSE loss functions were used for training. All combinations of the following architectures and parameter values were tested: $\{$ time\_steps $=[5,10,13,15,20]$, num\_layers $($ Conv1D $)=$ $[2,3,4,5]$, kernel\_sizes $=[3,5,10,12,15]$, filters $=[10,16, 32, 64]$, activations = ['swish', 'gelu', 'relu', 'tanh', 'sigmoid' $]\},\{$ learning rates $=[0.000025, 0.00015$, $0.0005,0.001]$, dropout rates $=[0.03,0.06,0.09,0.13]\}$. During the model selection, all models are trained for 75 epochs with early stopping with a patience of 10. The best TDNN model from our experiments uses time\_steps $=12$ for the data reshaping and has 3 Conv1D layers with tanh activations, filters $=12$, learning rate $=$ $2.5 \mathrm{e}-05$, dropout rate $=0.03$, and kernel\_size $=3$. It was trained for 200 epochs with a patience of 30 for the early stopping criterion, which was triggered after 175 epochs. As shown in Figure 23, this model's validation error fluctuates significantly throughout the training process, while its training error steadily decreases with minimal fluctuation. This divergence between the training and validation errors clearly indicates overfitting, where the model has learned to fit the noise in the training data rather than generalizing to unseen data. The large spikes in the validation error suggest that the model is not consistently performing well on the validation set, further emphasizing the lack of generalization. Specifically, overfitting is evident from several key signs in the plot: (1) the validation loss shows considerable variance with sharp increases at various points, indicating the model's sensitivity to the validation data; (2) the training loss decreases monotonically, suggesting that the model is continuing to learn patterns in the training data even as it becomes increasingly specific to that dataset, and (3) there is a clear divergence between the smooth, downward trend of the training loss and the erratic behavior of the validation loss, which reflects the model's inability to generalize effectively beyond the training set. These observations highlight the importance of adding regularization techniques such as l1/l2 penalty on the loss or larger dropout rate to prevent the model from fitting the training data too closely and to improve its performance on unseen data. As shown in Figure 24, the predicted C/K values for the TDNN model fit the actual values well, but the overpricing issue is apparent. Overall, as shown in Figure 28, it overprices $60.91 \%$, under-prices $18.75 \%$, and correctly prices $20.35 \%$ of calls. As shown in Figure 26, similar to the BS, MLP, and XGBoost models, the TDNN model does worse towards the end of the test set, but this is less noticeable. Also, similarly to the other models, the TDNN does slightly better on SPX than on NDX calls. As shown in Figure 27, this TDNN model does better than the BS models on all error metrics but slightly worse than MLP. 
\\
\section{RNN Model \& Self-attention Mechanism}
\myhypertarget{htag:Section10}{}
In time series regression tasks such as ours, where the objective is to model and predict temporal dependencies, Recurrent Neural Networks (RNNs) emerge as a natural choice due to their inherent ability to process sequences of data. Unlike traditional feedforward networks such as Multilayer Perceptrons (MLPs) or tree-based methods like XGBoost, which handle inputs in a static manner, RNNs are specifically designed to maintain and utilize information from previous time steps, thus capturing temporal patterns effectively. The core advantage of RNNs lies in their use of hidden states that are passed across time steps, enabling them to model sequential dependencies and handle variable-length input sequences, which MLPs and XGBoost cannot inherently manage without modification or preprocessing. However, despite their strengths, RNNs are not without drawbacks. One of the most critical challenges associated with training RNNs is the vanishing and exploding gradients problems, where gradients during backpropagation either diminish to near zero or escalate uncontrollably. This issue, as extensively discussed in Bengio et al. (1994) \cite{Bengio1994}, severely limits the network’s ability to learn long-term dependencies, often leading to suboptimal performance and difficulties during training.
\\

\subsection*{10.2 LSTM - 1997}
To mitigate these issues, advanced RNN variants such as Long Short-Term Memory (LSTM) networks and Gated Recurrent Units (GRUs) have been developed. LSTM networks, introduced by Hochreiter and Schmidhuber (1997) \cite{Hochreiter1997}, incorporate a gating mechanism that controls the flow of information, allowing the network to retain important information over extended periods and effectively combat the vanishing gradient problem. The key innovation in LSTMs is the memory cell state, which is modulated by input, forget, and output gates, ensuring that the network learns when to forget or retain information. The following pseudocode outlines a single pass through the LSTM algorithm:

\subsection*{Single Pass Through LSTM Algorithm}

\begin{enumerate}
    \item \textbf{Initialize} the hidden state $h_t$ and cell state $C_t$ for the current time step $t$.
    
    \item \textbf{Compute the Forget Gate:}
    \begin{enumerate}
        \item Concatenate the previous hidden state $h_{t-1}$ and current input $x_t$.
        \item Multiply the result by the weight matrix $W_f$ and add the bias $b_f$.
        \item Apply the sigmoid activation function to produce the forget gate output $f_t$.
    \end{enumerate}
    
    \item \textbf{Compute the Input Gate:}
    \begin{enumerate}
        \item Concatenate the previous hidden state $h_{t-1}$ and current input $x_t$.
        \item Multiply the result by the weight matrix $W_i$ and add the bias $b_i$.
        \item Apply the sigmoid activation function to produce the input gate output $i_t$.
    \end{enumerate}
    
    \item \textbf{Update the Cell State:}
    \begin{enumerate}
        \item Generate the candidate cell state $\tilde{C}_t$ using the previous hidden state $h_{t-1}$ and current input $x_t$.
        \item Multiply the result by the weight matrix $W_C$ and add the bias $b_C$.
        \item Apply the tanh activation function to obtain the candidate cell state.
        \item Update the cell state $C_t$ by combining the forget gate output $f_t$, the previous cell state $C_{t-1}$, and the input gate output $i_t$ with the candidate cell state $\tilde{C}_t$.
    \end{enumerate}
    
    \item \textbf{Compute the Output Gate:}
    \begin{enumerate}
        \item Concatenate the previous hidden state $h_{t-1}$ and current input $x_t$.
        \item Multiply the result by the weight matrix $W_o$ and add the bias $b_o$.
        \item Apply the sigmoid activation function to produce the output gate output $o_t$.
    \end{enumerate}
    
    \item \textbf{Compute the Hidden State:}
    \begin{enumerate}
        \item Apply the tanh activation function to the updated cell state $C_t$.
        \item Multiply the result by the output gate output $o_t$ to obtain the current hidden state $h_t$.
    \end{enumerate}
    
    \item \textbf{Return} the hidden state $h_t$ and the updated cell state $C_t$ as the output for time step $t$.
\end{enumerate}

The LSTM's gating mechanism can be mathematically stated as follows:

\begin{equation}
    \text{Forget Gate: } f_t = \sigma(W_f \cdot [h_{t-1}, x_t] + b_f)
\end{equation}

\begin{equation}
    \text{Input Gate: } i_t = \sigma(W_i \cdot [h_{t-1}, x_t] + b_i)
\end{equation}

\begin{equation}
    \text{Output Gate: } o_t = \sigma(W_o \cdot [h_{t-1}, x_t] + b_o)
\end{equation}

\begin{equation}
    \text{Cell State: } \tilde{C}_t = \tanh(W_C \cdot [h_{t-1}, x_t] + b_C)
\end{equation}

\begin{equation}
    \text{Cell Update: } C_t = f_t \cdot C_{t-1} + i_t \cdot \tilde{C}_t
\end{equation}

\begin{equation}
    \text{Hidden State: } h_t = o_t \cdot \tanh(C_t)
\end{equation}

\subsection*{10.2 GRU - 2014}

Gated Recurrent Units (GRUs), proposed by Cho et al. (2014) \cite{Cho2014} \cite{Chung2014}, offer a simpler alternative to LSTMs by streamlining the gating mechanism. Instead of having separate forget and input gates, GRUs combine these into a single update gate and introduce a reset gate to control the inclusion of past information. This architecture simplifies the model, making it more computationally efficient while retaining the ability to capture long-term dependencies. The following pseudocode outlines the steps involved in a single pass through the GRU:

\subsection*{Single Pass Through GRU Algorithm}
\begin{enumerate}
    \item \textbf{Initialize} the hidden state $h_t$ for the current time step $t$.

    \item \textbf{Compute the Reset Gate:}
    \begin{enumerate}
        \item Concatenate the previous hidden state $h_{t-1}$ and current input $x_t$.
        \item Multiply the result by the weight matrix $W_r$ and add the bias $b_r$.
        \item Apply the sigmoid activation function to produce the reset gate output $r_t$.
    \end{enumerate}
    
    \item \textbf{Compute the Update Gate:}
    \begin{enumerate}
        \item Concatenate the previous hidden state $h_{t-1}$ and current input $x_t$.
        \item Multiply the result by the weight matrix $W_z$ and add the bias $b_z$.
        \item Apply the sigmoid activation function to produce the update gate output $z_t$.
    \end{enumerate}
    
    \item \textbf{Compute the Candidate Hidden State:}
    \begin{enumerate}
        \item Apply the reset gate $r_t$ to the previous hidden state $h_{t-1}$ (element-wise multiplication).
        \item Concatenate the result with the current input $x_t$.
        \item Multiply the result by the weight matrix $W_h$ and add the bias $b_h$.
        \item Apply the tanh activation function to obtain the candidate hidden state $\tilde{h}_t$.
    \end{enumerate}
    
    \item \textbf{Compute the Final Hidden State:}
    \begin{enumerate}
        \item Combine the update gate output $z_t$ with the previous hidden state $h_{t-1}$ and the candidate hidden state $\tilde{h}_t$ using an element-wise multiplication and addition to obtain the final hidden state $h_t$.
    \end{enumerate}
    
    \item \textbf{Return} the final hidden state $h_t$ as the output for time step $t$.
\end{enumerate}

The GRU's gating mechanism can be mathematically described as follows:

\begin{equation}
    \text{Reset Gate: } r_t = \sigma(W_r \cdot [h_{t-1}, x_t] + b_r)
\end{equation}

\begin{equation}
    \text{Update Gate: } z_t = \sigma(W_z \cdot [h_{t-1}, x_t] + b_z)
\end{equation}

\begin{equation}
    \text{Candidate Hidden State: } \tilde{h}_t = \tanh(W_h \cdot [r_t \cdot h_{t-1}, x_t] + b_h)
\end{equation}

\begin{equation}
    \text{Final Hidden State: } h_t = z_t \cdot h_{t-1} + (1 - z_t) \cdot \tilde{h}_t
\end{equation}

In this formulation:

- $r_t$ is the reset gate, which determines how much of the past hidden state $h_{t-1}$ should be forgotten.
- $z_t$ is the update gate, which controls how much of the previous hidden state $h_{t-1}$ should be retained versus how much of the new candidate hidden state $\tilde{h}_t$ should be added.
- $\tilde{h}_t$ is the candidate hidden state, which is computed using the reset gate-modulated hidden state $r_t \cdot h_{t-1}$ and the current input $x_t$.
- $h_t$ is the final hidden state at time step $t$, which combines the old hidden state $h_{t-1}$ and the new candidate hidden state $\tilde{h}_t$ based on the values of the update gate $z_t$. These equations and the corresponding pseudocode outline the full operation of a GRU cell, which effectively manages information flow and retains long-term dependencies in the sequence data.

\subsection*{10.3 Model Selection}
Given the advantages of both LSTM and GRU architectures, we were motivated to test various combinations of these layers to determine the most effective architecture for our time series regression task. To prepare the data for RNN training, it was necessary to reshape the input data into sequences, ensuring that no future data was introduced into the past. This was accomplished using the following code:

\begin{verbatim}
# Function to reshape the data based on timesteps 
    # without introducing future data 
def reshape_data(data, timesteps):
    samples = data.shape[0] - timesteps
    reshaped_data=np.zeros((samples,timesteps,data.shape[1]))
    for i in range(samples):
        reshaped_data[i] = data[i:i + timesteps]
    return reshaped_data

# Reshape target variable y to match the number 
    # of samples in the reshaped X
def reshape_targets(y, timesteps):
    return y[timesteps:]
\end{verbatim}

\noindent We systematically tested all combinations of the following hyperparameters: \texttt{timesteps\_list = [3, 7, 12, 17, 25]}, \texttt{activations = ['tanh', 'relu', 'sigmoid']}, \texttt{neurons = [16, 32, 64, 128]}, and \texttt{learning\_rates = [0.000025, 0.00019, 0.00045, 0.0013, 0.0045, 0.021]}. Each model was trained over 100 epochs with early stopping and a patience of 20, using the Adam optimizer with a Mean Squared Error (MSE) loss function. After preliminary testing, we selected a dropout rate of 0.023, which showed consistent performance within a tested range of 1-35\%. The following architectures were evaluated:

\begin{verbatim}
# Model architectures
architectures = [
    # 3-layer architecture for LSTM only
    ('LSTM', 'LSTM', 'LSTM'),
    # 3-layer architecture for GRU only
    ('GRU', 'GRU', 'GRU'),
    # Specific 4-layer combinations for LSTM and GRU
    ('LSTM', 'GRU', 'LSTM', 'GRU'),
    ('LSTM', 'GRU', 'GRU', 'LSTM'),
    ('GRU', 'LSTM', 'GRU', 'LSTM'),
    ('GRU', 'LSTM', 'LSTM', 'GRU'),
    ('LSTM', 'LSTM', 'GRU', 'GRU'),
    ('GRU', 'GRU', 'LSTM', 'LSTM'), 
    # Specific 5-layer combinations for LSTM and GRU
    ('LSTM', 'GRU', 'LSTM', 'GRU', 'LSTM'),
    ('GRU', 'LSTM', 'GRU', 'LSTM', 'GRU') ]
\end{verbatim}

\noindent The best-performing architecture from our experiments is a hybrid recurrent neural network (RNN) combining both Long Short-Term Memory (LSTM) and Gated Recurrent Unit (GRU) layers, specifically configured in the sequence \texttt{('LSTM', 'GRU', 'LSTM', 'GRU', 'LSTM')}. This architecture is selected by testing all combinations of the parameters and architectures as described above. The specific hyperparameter values used are: \texttt{timestep = 12}, \texttt{activation = tanh}, \texttt{neurons = 32}, \texttt{learning\_rate = 0.000097}, and \texttt{dropout\_rate = 0.023}. The activation function across all layers was \texttt{tanh}, which is known for its smooth gradient properties, thus helping to mitigate vanishing gradient issues that can occur during backpropagation through time. Each recurrent layer was composed of 32 neurons, striking a balance between model complexity and computational efficiency. Following the model selection stage, we trained the best model for 250 epochs with early stopping and a patience of 30 epochs. This approach allowed the network to reach its optimal performance without overfitting, as the early stopping mechanism halted training once the validation loss ceased to improve significantly. The combination of LSTM and GRU layers in this architecture seems to provide the model with an enhanced capacity to learn and retain both short-term and long-term dependencies. This model does better than the BS, MLP, and TDNN on all error metrics. Despite the already impressive results of this model, we discuss adding a simple self-attention mechanism in the following section before comparing it fully to all models. 

\subsection*{10.4 Attention Mechanism \& Results}
We incorporate attention within the best-performing LSTM/GRU RNN architecture to enhance the model's ability to capture temporal dependencies and important patterns in the data. Our attention mechanism is inspired by the scaled dot-product attention used in transformers, tailored to suit RNNs \cite{Vaswani2017}. \noindent In this implementation, \( d_a \) represents the attention dimensionality, which is the number of features used in the attention mechanism equivalent to the 'neurons' parameter in our code. \noindent Given an input sequence represented by the hidden states $\mathbf{H} \in \mathbb{R}^{T \times d_a}$, where $T$ is the sequence length and $d_a$ is the dimensionality of the hidden states, the attention mechanism first computes query $\mathbf{Q}$, key $\mathbf{K}$, and value $\mathbf{V}$ matrices as linear transformations of $\mathbf{H}$:

\begin{equation}
\mathbf{Q} = \mathbf{H} \mathbf{W}_Q, \quad \mathbf{K} = \mathbf{H} \mathbf{W}_K, \quad \mathbf{V} = \mathbf{H} \mathbf{W}_V
\end{equation}

\noindent where $\mathbf{W}_Q, \mathbf{W}_K, \mathbf{W}_V \in \mathbb{R}^{d_a \times d_a}$ are learnable weight matrices, with $d_a = 32$ to match the number of neurons per layer in our best RNN configuration. The attention scores are computed using the scaled dot-product, followed by the application of the softmax activation function to ensure the scores are normalized:

\begin{equation}
\mathbf{A} = \text{softmax}\left(\frac{\mathbf{Q} \mathbf{K}^\top}{\sqrt{d_a}}\right)
\end{equation}

\noindent The softmax function is defined as:

\begin{equation}
\text{softmax}(z_i) = \frac{e^{z_i}}{\sum_{j} e^{z_j}}
\end{equation}

\noindent This step ensures that the attention scores sum to one, allowing them to effectively weigh the contributions of different input positions. The resulting attention matrix $\mathbf{A}$ is applied to the value matrix to produce the attention output:

\begin{equation}
\mathbf{O} = \mathbf{A} \mathbf{V}
\end{equation}

\noindent In our implementation, the attention mechanism is integrated into each RNN layer, allowing it to dynamically focus on salient features across different time steps. After obtaining the attention output, we concatenate it with the original RNN output, resulting in a more informative feature representation:

\begin{equation}
\mathbf{O}_{\text{concat}} = \text{concatenate}([\mathbf{O}, \mathbf{H}])
\end{equation}

\noindent To further enhance the model's performance, we apply the `tanh` activation function after each RNN layer:

\begin{equation}
\tanh(z) = \frac{e^{z} - e^{-z}}{e^{z} + e^{-z}}
\end{equation}

\noindent This activation function helps stabilize the learning process by maintaining gradients within a manageable range. Additionally, a dropout layer is applied after each RNN and attention mechanism to prevent overfitting and improve generalization. By dynamically weighting these features, the attention mechanism helps the model to capture nuanced market dynamics and improve generalization to unseen data \cite{Bahdanau2014, Luong2015}. In our experiments, the aforementioned best model enhanced with this simple self-attention mechanism performed very well and did better than all other models on all error metrics. This model was trained for 100 epochs with early stopping and patience of 30. As shown in Figure 33, this model's validation error fluctuates a lot at the start and throughout the 80 epochs before early stopping is triggered. Its training error almost does not fluctuate at all and just decreases quickly. As shown in Figure 31, the predicted C/K values for this model fit the actual values very well overall, and they provide the tightest fit around the true values of all models. Overall, it overprices 17.94\%, under-prices 34.01\%, and correctly prices 48.05\% of calls. As shown in Figure 32, similar to the other models, the best RNN does worse towards the end of the test set, but this is not as noticeable as for the other models. Also, similarly to the other models, it does slightly better on SPX than on NDX calls.

\section{KAN Model \& Kolmogorov-Arnold Representation Theorem}
\myhypertarget{htag:Section11}{}
Kolmogorov-Arnold Networks (KANs) represent an innovative neural network architecture that leverages the Kolmogorov-Arnold Representation Theorem to approximate multivariate functions through a systematic composition of univariate functions. This approach, as outlined in a recent study by Liu et al. \cite{Liu2024}, marks a significant departure from traditional feedforward neural networks and aims to leverage the strengths of MLPs while alleviating some of their weaknesses. Specifically, Kolmogorov-Arnold Networks represent an integration of splines and multilayer perceptrons, utilizing the strengths of both while mitigating their respective limitations \cite{Liu2024}. Splines are highly effective for approximating low-dimensional functions, allowing for precise local adjustments and the flexibility to adapt across different resolutions \cite{Liu2024}. However, they face significant challenges with the curse of dimensionality (COD), as they struggle to leverage compositional structures effectively \cite{Liu2024}. In contrast, MLPs are better equipped to handle the COD due to their inherent feature-learning capabilities \cite{Liu2024}. Nevertheless, MLPs lack the precision of splines in low-dimensional scenarios because of their inefficiency in optimizing univariate functions \cite{Liu2024}. As explained by Liu et al. in their study, for a model to accurately learn a function, it must be capable of 'both capturing the compositional structure (external degrees of freedom) and accurately approximating univariate functions (internal degrees of freedom)'. KANs achieve this by combining the advantages of both MLPs and splines—leveraging MLPs externally for feature learning and utilizing splines internally for optimal univariate function representation. This unique structure enables KANs to efficiently learn complex features, akin to MLPs, while simultaneously optimizing these features with the accuracy typical of spline-based methods \cite{Liu2024}.
\\

Instead of relying on fixed activation functions, KANs employ layers based on splines. In traditional feedforward neural networks, nonlinearities are typically introduced through activation functions like ReLU or sigmoid, applied element-wise to the outputs of linear transformations with learned weights and biases. In contrast, KANs capture these nonlinearities by constructing layers that apply splines with learnable coefficients to the inputs. Training KANs involves optimizing the constants in the splines and KAN layers are integrated into the network architecture in a manner similar to standard dense layers. In our implementation of KANs, we use orthogonal polynomials to construct these splines, which differs from the original approach that uses B-splines \cite{Liu2024}. Additionally, we employ dropout regularization to mitigate overfitting. In contrast, Liu et al. developed additional loss terms based on the l1 penalty on the splines' coefficients, an entropy regularization term, and performed targeted pruning of nodes to achieve sparsification. We anticipate that adding this targeted pruning mechanism instead of our dropout regularization would significantly improve our KANs. This is left for future work along with testing other modifications. Since all of the transformations are differentiable, KANs can be trained in the same way as MLPs via backpropagation using stochastic gradient descent or its variants, such as the Adam optimizer, which is used in this paper with the MSE loss function. 

\subsection*{11.1 Kolmogorov-Arnold Representation Theorem}
The Kolmogorov-Arnold Representation Theorem (KART) provides the theoretical foundation for KANs. It states that any continuous multivariate function \( f : [0, 1]^n \rightarrow \mathbb{R} \) can be represented as a finite composition of continuous univariate functions and addition. Formally, for any continuous function \( f(x_1, x_2, \dots, x_n) \), there exist continuous functions \( \phi_i : \mathbb{R} \to \mathbb{R} \) and \( \psi_{ij} : [0,1] \to \mathbb{R} \) such that:

\begin{equation}
f(x_1, x_2, \dots, x_n) = \sum_{i=1}^{2n+1} \phi_i \left( \sum_{j=1}^{n} \psi_{ij}(x_j) \right)
\end{equation}
\\

This theorem underscores the ability of KANs to approximate any continuous multivariate function by systematically constructing the network's architecture using the specified univariate transformations. Additionally, one of the key advantages of KANs lies in their inherent interpretability, which is particularly valuable in the domain of scientific machine learning, as highlighted by Liu et al. in their paper. Unlike traditional neural networks that often operate as "black boxes," KANs maintain a clear mathematical structure that aligns closely with the functional forms encountered in various scientific disciplines \cite{Liu2024}. This transparency allows researchers to better understand and trust the model's predictions, as each layer and transformation within a KAN has a clear and interpretable role. This makes KANs an attractive choice for applications where understanding the underlying relationships in the data is as important as the predictive performance itself \cite{Liu2024}. However, this aspect of KAN is not explored in our study and is left for future work. We refer the reader to the original paper \cite{Liu2024} for more information about the interpretability of KANs. Moreover, in their paper, Liu et al. demonstrate that the statement of the KART presented above corresponds to a composition of 2 KAN layers \cite{Liu2024}, which helps with understanding how more of these layers can be stacked to create deeper KANs. 

\subsection*{11.2 KAN as a Feedforward Neural Network}
Unlike the original Kolmogorov-Arnold Networks (KANs) as presented in \cite{Liu2024}, which utilize B-splines with an adaptive grid, our implementation uses simpler splines constructed from orthogonal polynomial bases without adaptive grid adjustment. Additionally, we do not employ residual activation functions at all and we use tanh to normalize the input before each layer's polynomial transformation which are two more differences from the original. Our KAN implementation also includes dropout for regularization (instead of the l1 norm and entropy-based loss on the spline coefficients as done in \cite{Liu2024}). We test four different types of orthogonal polynomials as alternatives for constructing splines: Legendre, Chebyshev of the 2nd kind, Laguerre, and Bessel. This decision was motivated by the powerful approximation properties of these polynomials but there are many other alternatives for the spline construction in KAN layers such as using wavelets \cite{BozChen2024} or Fourier coefficients \cite{XuChen2024} \cite{GN2024}. This description follows the detailed derivation provided in the original work \cite{Liu2024}, adapted to account for the inclusion of the \(\tanh\) normalization, orthogonal polynomial splines, and dropout layers in our modified KAN model. To get a sense of the network architecture as a whole, we start with an array specifying the number of nodes in the layers of the computational graph of the KAN (as done in \cite{Liu2024}). The computational graph has L+1 layers as the input is counted separately. 

\begin{equation}
[n_0, \dots, n_{L-1}, n_{L}],
\end{equation}

\noindent Here, $n_{i}$ is the number of nodes in the $i$-th layer of the computational graph, with $n_{0}$ being the (zeroth) input layer. Next, we specify the input and output dimensions of each layer of the KAN and choose the polynomials to apply at each layer with the degree to go up to in the spline:

\begin{equation}
[\mathbf{\Phi}^{(0)}: (n_0, n_1); \dots ; \mathbf{\Phi}^{(L-1)}:(n_{L-1}, n_{L})],
\end{equation}

\begin{equation} \left[ \text{Layer } 1: P^{(1)}(x), D_1; \ \dots \ ; \ \text{Layer } L: P^{(L)}(x), D_L \right], \end{equation}

\noindent Hence, the input and output dimensions of the $(j)$-th KAN layer are $n_{j-1}$ and $n_{j}$ respectively and $j$ ranges from 1 to $L$. However, we index the KAN layers such that $\mathbf{\Phi}^{(l-1)}$ is the $l$-th KAN layer with $l$ ranging from 1 to $L$. In this study, we restrict our experiments to KAN architectures with the same choice of orthogonal polynomials used for the splines in each KAN layer. We implore the reader to note that we refer to both a 'layer' of the computational graph and KAN 'layers' but these are different. Namely, the $l$-th KAN layer goes between the $(l-1)$-th and $l$-th layers of the computational graph. There are $n_{l-1} \times n_{l}$ activation functions between layers $(l-1)$ and $l$ of the computational graph \cite{Liu2024}, which corresponds to the $l$-th KAN layer. Specifically, the activation function connecting neuron $p$ in layer $(l-1)$ to neuron $q$ in layer $l$ (of the computational graph) is denoted by:

\begin{equation}
\phi^{(l-1)}_{q,p} \quad \textit{for} \quad l = 1, \dots, L ; \quad p = 1, \dots, n_{l-1} ; \quad q = 1, \dots, n_{l}.
\end{equation}

\noindent The pre-activation value at neuron $p$ in layer $(l-1)$ (of the computational graph) is $x_p^{(l-1)}$, while the post-activation value is:

\begin{equation}
\tilde{x}^{(l-1)}_{q,p} \equiv \phi^{(l-1)}_{q,p}\left( x_p^{(l-1)} \right),
\end{equation}

\noindent which represents the output of the activation function applied to the pre-activated value. For the $q$-th neuron in layer $l$ of the computational graph, the overall activation value is then the sum of all incoming post-activations:

\begin{equation}
x_q^{(l)} = \sum_{p=1}^{n_{l-1}} \tilde{x}^{(l-1)}_{q,p} = \sum_{p=1}^{n_{l-1}} \phi^{(l-1)}_{q,p}\left( x_p^{(l-1)} \right), \quad q = 1, \dots, n_{l}.
\end{equation}

\noindent For \(n_{l-1}\)-dimensional inputs and \(n_{l}\)-dimensional outputs, the $l$-th KAN layer in our implementation can be defined as applying the following transformation to the data:

\begin{equation}
\mathbf{\Phi}^{(l-1)} = \begin{bmatrix} \phi^{(l-1)}_{q,p} \end{bmatrix}, \quad p = 1, 2, \dots, n_{l-1}, \quad q = 1, 2, \dots, n_{l},
\end{equation}

\noindent where each $\phi^{(l-1)}_{q,p}$ is a composition of \(\tanh\) and the spline of orthogonal polynomials as outlined below:

\begin{equation}
\phi^{(l-1)}_{q,p}(x_p^{(l-1)}) = W^{(l)}_{q,p} \cdot \sum_{n=0}^{D_l} \tilde{W}^{(l)}_{q, p, n} \cdot P^{(l)}_n\left( \tanh(x_p^{(l-1)}) \right),
\end{equation}

\noindent where:

\begin{itemize}
    \item $x_p^{(l-1)}$ is the pre-activation value of the $p$-th neuron in layer $l$ of the computational graph.
    \item $P^{(l)}_n(\tanh(x_p^{(l-1)}))$ is the orthogonal polynomial chosen for layer (of the KAN) $l$ of degree $n$ applied to the $p$-th component of the $\tanh$-normalized input.
    \item $D_l$ is the largest degree of the polynomials used in layer $l$.
    \item $W_{q,p}^{(l)}$ and $\tilde{W}_{q, p, n}^{(l)}$ are the learnable weights.
\end{itemize}

\noindent Hence, for the $q$-th neuron in layer $l$, the overall activation value can then be written as: 
\begin{equation}
x_q^{(l)} = \sum_{p=1}^{n_{l-1}} W^{(l)}_{q,p} \cdot \sum_{n=0}^{D_l} \tilde{W}^{(l)}_{q, p, n} \cdot P^{(l)}_n\left( \tanh(x_p^{(l-1)}) \right) , \quad q = 1, \dots, n_{l}. 
\end{equation}

\noindent For an $L$-layer KAN, the total number of weights can be computed as follows: 
\begin{align}
\sum_{l=1}^{L} \left(\left(n_{l-1} \cdot n_{l}\right) + \left(n_{l-1} \cdot n_{l} \cdot \left(D_{l} + 1\right)\right)\right) = \sum_{l=1}^{L} \left(n_{l-1} \cdot n_{l} \cdot \left(D_{l} + 2 \right) \right)
\end{align}

\noindent The $\tanh$ function ensures that the input remains within the stable range $[-1, 1]$, which is critical for maintaining the stability and accuracy of polynomial computations, especially when higher-degree polynomials are used. We also add dropout after each KAN layer (excluding the first and last KAN layers). The dropout function is applied (after the KAN layer transformation) at the 2, ..., (L - 1) layers' outputs, randomly zeroing a fraction of the outputs during training. Here we denote \(\gamma\) percent dropout as \(\textbf{D}_{\gamma}(\cdot)\). We use the standard implementation of inverted dropout in PyTorch. The dropout operation $\mathbf{D}_{\gamma}$ can be represented as:

\begin{equation} \mathbf{D}_{\gamma}\left( \mathbf{z} \right) = \frac{\mathbf{m} \odot \mathbf{z}}{1 - \gamma}, \end{equation}

\noindent where $\mathbf{z}$ is the input vector to the dropout layer, $\mathbf{m}$ is a mask vector of the same dimension as $\mathbf{z}$, with each element $m_i$ independently sampled from a Bernoulli distribution:

\begin{equation} 
m_i \sim \text{Bernoulli}(1 - \gamma), 
\end{equation}

\noindent and $\odot$ denotes element-wise multiplication. During training, $\gamma$ is the dropout rate (e.g., $\gamma=0.3$ for 30\% dropout) and during testing, dropout is disabled. The final output of our KANs, representing the prediction \(\hat{y}\), can be written as:

\begin{equation}
    \hat{y} = \left(\mathbf{\Phi}^{(L-1)} \circ \textbf{D}_{\gamma} \circ \mathbf{\Phi}^{(L-2)} \circ \cdots \circ \textbf{D}_{\gamma} \circ \mathbf{\Phi}^{(1)} \circ \mathbf{\Phi}^{(0)}\right)(\mathbf{x}^{(0)}),
\end{equation}

\noindent where \(L\) is the total number of KAN layers in the network, and \(\mathbf{\Phi}^{(l-1)}\) is the KAN transformation applied by the \(l\)-th layer. Here, $\mathbf{x}^{(0)}$ is the input with $n_0$ components and $\hat{y} = \mathbf{x}^{(L)}$ is the output with $n_L$ components. 
\\

Our experiments for the KAN models are carried out using PyTorch, where we implement custom KAN layers to compute the output based on the recurrence relations of the chosen orthogonal polynomials. The coefficients $W_{q,p}^{(l)}$ for each layer are initialized using a normal distribution with a mean of 0 and a standard deviation that is inversely proportional to the sum of the input and output dimensions of that layer. Specifically, the weights $W_{q,p}^{(l)}$ are sampled as follows:

\begin{equation}
W^{(l)}_{q,p} \sim \mathcal{N}\left( 0, (\sqrt{\frac{5}{ (n_{l-1} + n_{l})}})^2 \right),
\end{equation}

\noindent Additionally, we initialize the spline coefficients sampled from a normal distribution centered around 0.1 with a standard deviation inversely proportional to the degree: 
\begin{equation}
\tilde{W}^{(l)}_{q, p, n} \sim \mathcal{N}\left( 0.1, (\frac{1}{ (D_l+1)})^2 \right),
\end{equation}

\subsection*{11.4 Recursive Definitions of Orthogonal Polynomials}
Figure 39 shows plots of the first seven polynomials for the various families of orthogonal polynomials in our study. In our implementation, we employed the following recursive definitions:

\textbf{Chebyshev Polynomials of the Second Kind}
\begin{align}
C_0(x) &= 1 \\
C_1(x) &= 2x \\
C_{n+1}(x) &= 2xC_n(x) - C_{n-1}(x)
\end{align}

\textbf{Legendre Polynomials}
\begin{align}
Lg_0(x) &= 1 \\
Lg_1(x) &= x \\
(n+1)Lg_{n+1}(x) &= (2n + 1)xLg_n(x) - nLg_{n-1}(x)
\end{align}

\textbf{Bessel Polynomials}
\begin{align}
B_0(x) &= 1 \\
B_1(x) &= x + 1 \\
B_n(x) &= (2n - 1)xB_{n-1}(x) + B_{n-2}(x)
\end{align}

\textbf{Laguerre Polynomials}
\begin{align}
L_0(x) &= 1 \\
L_1(x) &= 1 - x \\
L_{k+1}(x) &= \frac{(2k + 1 - x) L_k(x) - k L_{k-1}(x)}{k + 1}
\end{align}

\noindent These recursive formulations were the simplest to implement the KAN layers. However, using these recursive definitions is not the most computationally efficient method, as using some of the series or closed-form expressions for the polynomials could be quicker. Moreover, it would be prudent to test the alternative of using min-max normalization instead of tanh to keep the inputs in the [-1, 1] range. Below is an example of how a Legendre KAN layer is implemented in our framework:

\begin{verbatim}
class LegendreKANLayer(nn.Module):
    def __init__(self, input_dim, output_dim, degree):
        super(LegendreKANLayer, self).__init__()
        self.inputdim = input_dim    # n_{l-1}
        self.outdim = output_dim   # n_{l}
        self.degree = degree      # D_l

        # initialize W^{(l)}_{q,p}
        self.W = nn.Parameter(torch.empty(self.outdim, \
        self.inputdim))
        std_W = torch.sqrt(5.0 / (self.inputdim + self.outdim))
        nn.init.normal_(self.W, mean=0.0, std=std_W)

        # initialize \tilde{W}^{(l)}_{q,p,n}
        self.tilde_W = nn.Parameter(torch.empty(self.outdim, \
                        self.inputdim, self.degree + 1))
        std_tilde_W = 1.0 / (self.degree + 1)
        nn.init.normal_(self.tilde_W, mean=0.1, std=std_tilde_W)

    def forward(self, x):
        # Reshape input to (batch_size, n_{l-1})
        x = x.view(-1, self.inputdim)
        # Apply tanh to get within [-1, 1]
        x = torch.tanh(x)
        batch_size = x.size(0)

        # compute Legendre polynomials up to specified degree
        legendre_list = [ ]
        legendre_0 = torch.ones(batch_size, \
                        self.inputdim, device=x.device)
        legendre_list.append(legendre_0)
                              
        if self.degree > 0:
            legendre_1 = x_tanh
            legendre_list.append(legendre_1)
        for n in range(2, self.degree + 1):
            legendre_n = ((2 * n - 1) / n) * x_tanh \
                        * legendre_list[n - 1] \
                         - ((n - 1) / n) * \
                         legendre_list[n - 2]
            legendre_list.append(legendre_n)

        # stack polynomials: Shape (batch_size, \
                            input_dim, degree + 1)
        legendre = torch.stack(legendre_list, dim=2) 

  #s_{b,q,p} = sum_{n=0}^{D_l} \tilde{W}_{q,p,n} * P_n(x_p)
        # legendre: (batch_size, input_dim, degree + 1)
        # tilde_W: (output_dim, input_dim, degree + 1)
        # result s: (batch_size, output_dim, input_dim)
        s = torch.einsum('bpn,qpn->bqp', \
                        legendre, self.tilde_W)

        # multiply by W^{(l)}_{q,p} and sum over p
        # W: (output_dim, input_dim)
        # s: (batch_size, output_dim, input_dim)
        # y: (batch_size, output_dim)
        y = torch.einsum('bqp,qp->bq', s, self.W) 

        return y
\end{verbatim}

\noindent This custom layer generates the KAN transformation for each layer based on the Legendre polynomials, while similar classes handle the other orthogonal polynomials. In \cite{Liu2024}, the network uses B-splines for the activation functions with additional grid extension mechanisms to adjust the grid dynamically during training. However, in our implementation, we rely purely on the intrinsic properties of the orthogonal polynomials, which provide stability and efficiency without the need for grid adjustments. In future work, it would be prudent to test whether combining layers of different polynomials increases performance. Also, from our experiments, it is clear that our KANs would benefit significantly from more regularization since dropout alone is not enough. We anticipate that the targeted pruning of nodes mechanism used in the original paper should work very well. 

\subsection*{11.3 KAN Best Model Selection and Testing}
In our study, an extensive model selection process was conducted to identify the best KAN architecture for our task. We systematically varied several hyperparameters, including the type of orthogonal polynomial used in the KAN layers, the number of neurons per layer (8, 16, 32, 64), the number of KAN layers (2 to 4), the learning rates (0.000055 to 0.02), and the degrees of the polynomials (1 to 7). The KANs were trained using the Adam optimizer with a Mean Squared Error (MSE) loss function, as with the other models. In the best model selection, each model was trained for 75 epochs, with early stopping implemented (patience of 25 epochs) to prevent overfitting. We used four types of KAN layers (Legendre, Chebyshev of the Second Kind, Bessel, and Laguerre), systematically testing all combinations of polynomial degrees and layers. The best model was identified based on the validation error. Our systematic exploration of these combinations allows us to evaluate the effectiveness of different configurations to identify the optimal model architecture. In our experiments, the best KAN model had the following configuration: ['kan\_layer': Cheby2KANLayer, 'neurons': 16, 'layers': 3, 'learning\_rate': 0.0059, 'degree\_combination': [2, 5, 4] ] and performed very well. In fact, it does better than the BS, MLP, and TDNN on all error metrics, as shown in Figure 27. This model was trained for 100 epochs with early stopping and patience of 30 and a dropout of 5\%. As shown in Figure 37, this model’s validation error fluctuates a lot but steadily decreases over time, and its training error is much more stable and also goes down steadily over epochs. As shown in Figure 35, the predicted C/K values for this model fit the actual values very well. Overall, it overprices 21.42\%, under-prices 44.06\%, and correctly prices 34.52\% of calls. As shown in Figure 36, similarly to the MLP, TDNN, XGBoost, and RNN models, this model does worse towards the end of the test set. 

\section{Evaluation \& Results}
\myhypertarget{htag:Section12}{}
As can be seen in Figure 27, which is a table of error metrics by model, the best-performing model across all error metrics was the LSM-GRU hybrid RNN model with attention. However, all the other models also did significantly better than the B-S model. Although TDNN did significantly better than BS, it was the second worst model, and even our simple MLP did slightly better on all error metrics. This is likely because we did not implement a full hierarchy of varying temporal convolutions as intended in the original TDNN architecture. As can be seen in Figure 28, which is a table of over/under/correctly priced proportions, the RNN and the TDNN models have the lowest and largest proportions of overpriced calls, respectively. Moreover, the RNN and the B-S models have the largest and lowest proportions of correctly priced calls, respectively. Finally, the TDNN and MLP models have the lowest and largest proportions of under-priced calls, respectively. Interestingly, although the RNN model is the best, it has more of an issue with underpricing than the TDNN and BS models. As shown in Figure 29, all models perform better on SPX than NDX calls in terms of correctly priced $\%$. The difference in performance for the two tickers is smallest for the KAN model. The RNN, XGBoost, TDNN, and MLP models all overprice more on NDX than on SPX calls, but the opposite is true for the BS and KAN models. The RNN, TDNN, and XGBoost models all underprice more on SPX than on NDX calls, but the opposite is true for the BS, MLP, and KAN models. 
\\

As expected, Figure 30 shows that all models do better on ITM than on ATM calls (in terms of $\%$ correctly priced). But, unexpectedly, all models do better on ATM than on OTM calls, which may be because the range around the value of 1 for the ATM category is too wide. Also, all models have larger \% overpriced than under-priced for the ATM moneyness category. The MLP and B-S models display a higher percentage of overpriced than underpriced options for the ITM category, whereas the other models exhibit the opposite trend, with a greater percentage of underpriced than overpriced options for ITM calls. Furthermore, all models aside from the MLP have a higher percentage of overpriced than underpriced options for OTM calls. Given that some of the errors between the models are complementary, exhibiting opposite percentages of overpriced and underpriced options across certain moneyness categories, it may be beneficial to explore ensembling approaches to leverage these complementary strengths.
\\

\section{Future Work}
\myhypertarget{htag:Section13}{}
To advance this research, we plan to finalize the implementation of the Time-Delay Neural Network (TDNN) by incorporating a carefully designed hierarchy of temporal convolutions. Following this, we will integrate several enhancements into the MLP, KAN, and RNN models, including the application of alternative regularization techniques, the implementation of batch and layer normalization, and the evaluation of alternative activation functions such as Swish and GELU. To further improve the KAN and MLP models, we will explore the impact of integrating LSTM or Gated Recurrent Unit (GRU) layers and attention mechanisms. Additionally, we will implement the explicit formula definitions for orthogonal polynomials, which should enhance the computational efficiency of the KANs. We also plan to experiment with multi-head attention as an alternative to the current self-attention mechanism employed in our models. Another aspect of this study we will tweak further is the margin for correctly priced options and the ATM range. We want to test whether narrowing both may yield a better comparison across models. After these adjustments, we intend to expand the feature set by incorporating volume, skewness, and kurtosis and adapt all models to account for put options in addition to call options by leveraging put-call parity. Moreover, we hope to extend the analysis to include other financial indices and investigate the potential benefits of applying smoothing techniques to the historical volatility and interest rate estimates. We also plan to compare our models to Heston stochastic volatility and other extensions of the BS model by discretizing the SDEs and pricing our call options via Monte Carlo simulation, which facilitates the comparison of our existing models to an alternative pricing method that is widely used in the industry and academia. Another direction we want to explore is to check which models capture the volatility smile best and compare this to the BS model's inability to capture it. Moreover, we want to investigate the specific polynomials and coefficients used in the KAN model for the final predictions, leveraging the interpretability of KANs. 
\\

In addition to the improvements discussed, we plan to explore alternative supervised learning model architectures for this task. One promising direction involves the use of selective state space models (SSMs), particularly Mamba \cite{Gu2023}, and its variants—Mamba-2 \cite{Dao2024}, DyGMamba \cite{Ding2024}, and Mamba-2-Hybrid \cite{Wal2024}, which have been shown to perform well in time series forecasting tasks. Mamba is a novel architecture designed to capture temporal dependencies in data. It integrates selective state space models into its framework, allowing the model to dynamically propagate or forget information based on the sequence length dimension and input content, leading to significant improvements in efficiency and performance, especially on long sequences \cite{Gu2023}. Building on Mamba, Mamba-2 introduces a structured state space model (SSM) that unifies the theoretical connections between SSMs and variants of attention mechanisms. This refinement allows Mamba-2 to generalize well across multiple tasks while achieving linear scaling in sequence length. Furthermore, Mamba-2’s core architecture has been shown to outperform Transformers of equivalent size and complexity, with significant efficiency gains \cite{Dao2024}. DyGMamba extends this by incorporating dynamic temporal graphs into the SSM framework. This allows for efficient modeling of long-term temporal dependencies on continuous-time dynamic graphs, effectively capturing intricate temporal patterns that emerge over extended periods. DyGMamba achieves state-of-the-art performance on temporal graph learning tasks, combining computational efficiency with powerful temporal representation learning \cite{Ding2024}. 
\\

These models have demonstrated a 5x increase in throughput over Transformers while maintaining state-of-the-art performance across various modalities, including language, audio, and genomics \cite{Gu2023}. However, to the best of our knowledge, they have not been explored for financial time series in the context of option pricing, although other state space models like Hidden-Markov models (HMMs) have been. Mamba and its variants offer promising avenues for future research, and we plan to rigorously evaluate their performance on our task, comparing them with the existing models to determine their effectiveness in capturing complex financial time-series patterns. Furthermore, we will explore Physics-Informed Neural Networks (PINNs), given that the option pricing problem can be framed as learning the solution to a partial differential equation (PDE) from noisy data. This approach could provide additional insights and improved accuracy in modeling financial derivatives. Also, we plan to test CatBoost as an alternative gradient-boosting decision tree model since it has been shown to outperform XGBoost on time series forecasting and tasks with tabular data \cite{Shmuel2024}. Finally, we plan to test the performance of a KAN-Capsule-Net as a novel architecture for this task. This is an architecture inspired by Hinton's CapsNet but with KAN layers with orthogonal polynomials instead of MLP blocks. Capsule Networks (CapsNets) is a type of neural network architecture aiming to better capture spatial/temporal hierarchies and relationships between features using "capsules" composed of small groups of neurons. Hinton et al. proposed that the key advantage of CapsNets is their ability to dynamically route information between capsules, allowing the network to model part-whole relationships more effectively than traditional ANNs or convolutional neural networks (CNNs) \cite{Sabour2017}.

\section*{Appendix: Best KAN Model Parameter Count \& Equations}
\myhypertarget{htag:Appendix}{}

\noindent In our experiments, the best KAN has the following configuration from the code: \\
\verb|['kan_layer': Cheby2KANLayer, 'neurons': 16, 'layers': 3,| \\
\verb|'learning_rate': 0.0059, 'degree_combination': [2, 5, 4]]|. \\

\noindent This can also be specified as follows: 
\begin{equation}
[n_0 = 10, n_1 = 16, n_2 = 16, n_3 = 1],
\end{equation}

\begin{equation}
[\mathbf{\Phi}^{(0)}: (10, 16); \mathbf{\Phi}^{(1)}: (16, 16);\mathbf{\Phi}^{(2)}: (16, 1)],
\end{equation}

\begin{equation} \left[ C^{(i)}(x), \forall \mathbf{i}; D_1 = 2, D_2 = 5, D_3 = 4 \right], \end{equation}

\noindent \textbf{Parameter Count:}

\noindent For each KAN layer \( l \):
\begin{itemize}
    \item \textbf{Polynomial Spline Weights:} $n_{l-1} \times n_{l} \times (D_l + 1)$
\end{itemize}

\begin{itemize}
    \item \textbf{Outer Sum Weights:} $n_{l-1} \times n_{l}$
\end{itemize}

\noindent - \textbf{Layer 1 (10 Input Neurons to 16 Output Neurons) with $D_1 = 2$:}
\begin{itemize}
    \item Spline Weights: \( 10 \times 16 \times (2 + 1) = 480 \) parameters.
    \item Outer Sum Weights: \( 10 \times 16  = 160 \) parameters.
\end{itemize}

\noindent - \textbf{Layer 2 (16 Neurons to 16 Output Neurons) with $D_2 = 5$:}
\begin{itemize}
    \item Spline Weights: \( 16 \times 16 \times (5 + 1) = 1536 \) parameters.
    \item Outer Sum Weights: \( 16 \times 16  = 256 \) parameters.
\end{itemize}

\noindent - \textbf{Layer 3 (16 Neurons to 1 Output Neuron) with $D_3 = 4$:}
\begin{itemize}
    \item Spline Weights: \( 16 \times 1 \times (4 + 1) = 80 \) parameters.
    \item Outer Sum Weights: \( 16 \times 1  = 16 \) parameters.
\end{itemize}

\noindent \textbf{Total Parameter Count for the KAN Model:} \( (480 + 160) + (1536 + 256) + (80 + 16) = 2528 \) parameters.
\\

\noindent \textbf{Comparison to a standard 3 hidden layer MLP with 16 Neurons per Layer:}
\begin{itemize}
\item Layer 1 (Input to First Hidden Layer): \( 10 \times 16 + 16 = 176 \) parameters.
\item Layer 2 (First to Second Hidden Layer): \( 16 \times 16 + 16 = 272 \) parameters.
\item Layer 3 (Second to Third Hidden Layer): \( 16 \times 16 + 16 = 272 \) parameters.
\item Layer 4 (Third Hidden Layer to Output): \( 16 \times 1 + 1 = 17 \) parameters.
\end{itemize}
Total for the MLP: \( 176 + 272 + 272 + 17 = 737 \) parameters.
\\

\noindent Therefore, our three-layer KAN model with Chebyshev polynomials has \( 2528 - 737 = 1791 \) more parameters compared to a standard three hidden layer MLP with the same number of neurons in each layer and the same input and output dimensions. For an $L$-layer KAN, the total number of parameters can be computed as follows: 
\begin{align}
\sum_{l=1}^{L} \left(\left(n_{l-1} \cdot n_{l}\right) + \left(n_{l-1} \cdot n_{l} \cdot \left(D_{l} + 1\right)\right)\right) = \sum_{l=1}^{L} \left(n_{l-1} \cdot n_{l} \cdot \left(D_{l} + 2 \right) \right)
\end{align}

\noindent The final output of our best KAN variant, representing the prediction \(\hat{y}\), can be written as:

\begin{equation}
    \hat{y} = \left( \mathbf{\Phi}^{(2)} \circ \textbf{D}_{\gamma} \circ \mathbf{\Phi}^{(1)} \circ \mathbf{\Phi}^{(0)} \right)(\mathbf{x}^{(0)}) ,
\end{equation}

\noindent Here, $\mathbf{x}^{(0)}$ is the input with $n_0 = 10$ components and $\hat{y} = \mathbf{x}^{(3)}$ which is a scalar since $n_3 = 1$. In the fllowing equations, $\mathbf{x}^{(1)}$ has 16 components, $\mathbf{x}^{(2)}$ and $\mathbf{\bar{x}}^{(2)}$ have 16 components, and $\mathbf{x}^{(3)}$ is a scalar. We represent $\mathbf{x}^{(2)}$ after dropout as $\mathbf{\bar{x}}^{(2)}$ and $\mathbf{m}$ is a mask vector with 16 components to be applied via element-wise multiplication with $\mathbf{\bar{x}}^{(2)}$. In our best KAN configuration, we use a dropout of 5\% meaning $\gamma$ = 0.05 in the equations below. 
\\

\begin{align}
\mathbf{x}^{(1)} = \sum_{p_1=1}^{n_0=10} W^{(1)}_{p_2,p_1} \sum_{k_1=0}^{D_1=2} \tilde{W}^{(1)}_{p_2,p_1,k_1} \cdot C^{(1)}_{k_1} \Big( \mathbf{tanh}(x^{(0)}_{p_1}) \Big)
\end{align}

\begin{align}
\mathbf{x}^{(2)} = \sum_{p_2=1}^{n_1=16} W^{(2)}_{p_3,p_2} \sum_{k_2=0}^{D_2=5} \tilde{W}^{(2)}_{p_3,p_2,k_2} \cdot C^{(2)}_{k_2} \Big( \mathbf{tanh}(x^{(1)}_{p_2}) \Big)
\end{align}

\begin{align}
\mathbf{\bar{x}}^{(2)} = \textbf{D}_{\gamma} \Big(\mathbf{x}^{(2)}\Big) = \frac{\mathbf{m}}{1 - \gamma} \odot \mathbf{x}^{(2)}
\end{align}

\begin{align}
\hat{y} = \mathbf{x}^{(3)} = \sum_{p_3=1}^{n_2=16} W^{(3)}_{1,p_3} \sum_{k_3=0}^{D_3=4} \tilde{W}^{(3)}_{1,p_3,k_3} \cdot C^{(3)}_{k_3} \Big( \mathbf{tanh}(\bar{x}^{(2)}_{p_3}) \Big)
\end{align}

\section*{Figures}
\myhypertarget{htag:Figures}{}

\begin{itemize}
  \item Figure 1: Risk-Free Rate vs Time
\end{itemize}

\begin{center}
\includegraphics[max width=\textwidth]{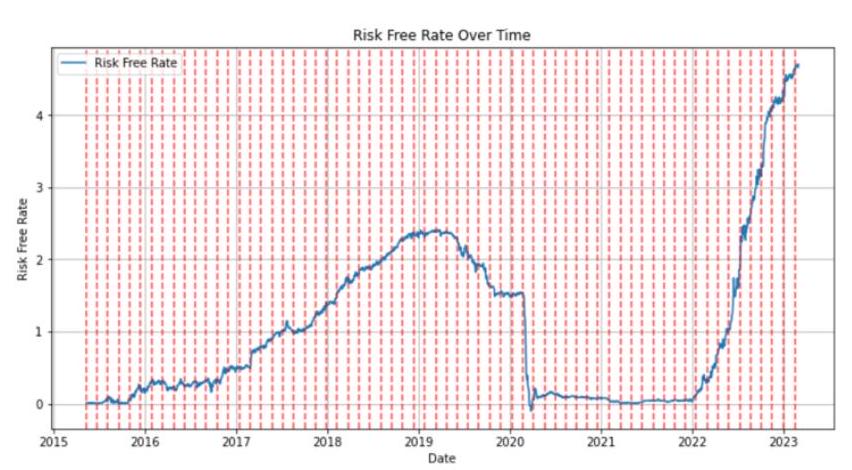}
\end{center}

\begin{itemize}
  \item Figure 2: Historic Volatilities by Ticker for Different Window Size
\end{itemize}

\begin{center}
\includegraphics[max width=\textwidth]{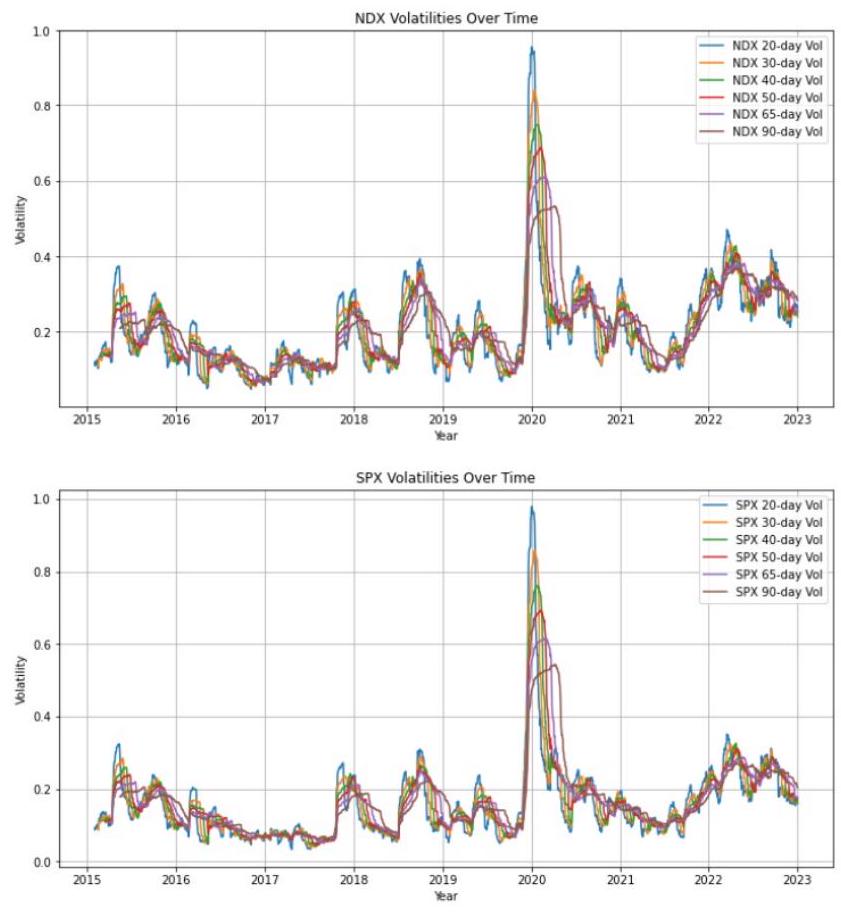}
\end{center}

\begin{itemize}
  \item Figure 3: Log Price Returns Distribution
\end{itemize}

\begin{center}
\includegraphics[max width=\textwidth]{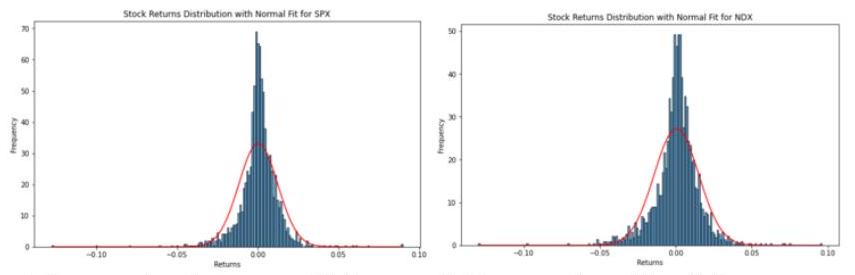}
\end{center}

\begin{itemize}
  \item Figure 4: Average Price and Proportion for Moneyness Category by Ticker
\end{itemize}

\begin{center}
\begin{tabular}{rrrr}
\multicolumn{4}{c}{Moneyness\_Category} \\
\hline
Average Price \$ & ATM (0.95-1.05) & 568.028975 & 132.819948 \\
 & ITM (>1.05) & 1596.283873 & 447.234934 \\
Proportion & OTM (<0.95) & 144.466523 & 30.375881 \\
 & ATM (0.95-1.05) & 0.318382 & 0.382268 \\
 & ITM (>1.05) & 0.247259 & 0.301388 \\
 & OTM (<0.95) & 0.434359 & 0.316343 \\
\end{tabular}
\end{center}

\begin{itemize}
  \item Figure 5: Average Price and Proportion for Moneyness Category by Time to Expiration
\end{itemize}

\begin{center}
\begin{tabular}{rrrrrr}
 & \begin{tabular}{r}
Expiry\_Category \\
Moneyness\_Category \\
\end{tabular} & $\mathbf{6 0 - 1 8 0}$ & $<60$ & $>180$ \\
\hline
Average Price \$ & ATM (0.95-1.05) & 264.520889 & 184.481992 & 700.696291 \\
 & ITM (>1.05) & 731.065316 & 819.329338 & 1304.859530 \\
Proportion & OTM (<0.95) & 57.828998 & 12.396391 & 259.054764 \\
 & ATM (0.95-1.05) & 0.110135 & 0.178742 & 0.065338 \\
 & ITM (>1.05) & 0.115554 & 0.096874 & 0.065193 \\
 & OTM (<0.95) & 0.110588 & 0.162889 & 0.094687 \\
\end{tabular}
\end{center}

\begin{itemize}
  \item Figure 6: Error Metrics for B-S Model with Different Volatility Estimates
\end{itemize}

\begin{center}
\begin{tabular}{|l|l|l|l|}
\hline
\begin{tabular}{l}
Yol. Windowys \\
Efror Metrics on \\
Teft-fet \\
\end{tabular} & MSE (6 d.p.) & RMSE (6 d.p.) & MAE (6 d.p.) \\
\hline
\begin{tabular}{l}
BS w. 20-day \\
window vol. \\
\end{tabular} & 0.003528 & 0.059396 & 0.022639 \\
\hline
\begin{tabular}{l}
BS w. 30-day \\
window vol. \\
\end{tabular} & 0.003451 & 0.058743 & 0.021894 \\
\hline
\begin{tabular}{l}
BS w. 40-day \\
window vol. \\
\end{tabular} & 0.003375 & 0.058091 & 0.020700 \\
\hline
\begin{tabular}{l}
BS w. 50-day \\
window vol. \\
\end{tabular} & 0.003306 & 0.057499 & 0.019714 \\
\hline
\begin{tabular}{l}
BS w. 65-day \\
window vol. \\
\end{tabular} & 0.003228 & 0.056814 & 0.019096 \\
\hline
\begin{tabular}{l}
BS w. 90-day \\
window vol. \\
\end{tabular} & 0.003142 & 0.056056 & 0.018677 \\
\hline
\end{tabular}
\end{center}

\begin{itemize}
  \item Figure 7: Predicted vs Actual C/K for Best B-S Model
\end{itemize}

\begin{center}
\includegraphics[max width=\textwidth]{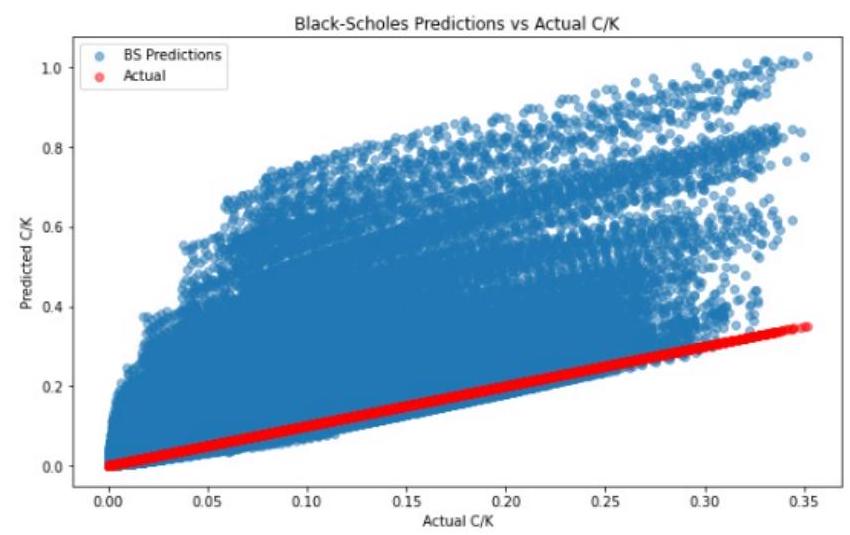}
\end{center}

\begin{itemize}
  \item Figure 8: Error Distribution for Best B-S Model
\end{itemize}

\begin{center}
\includegraphics[max width=\textwidth]{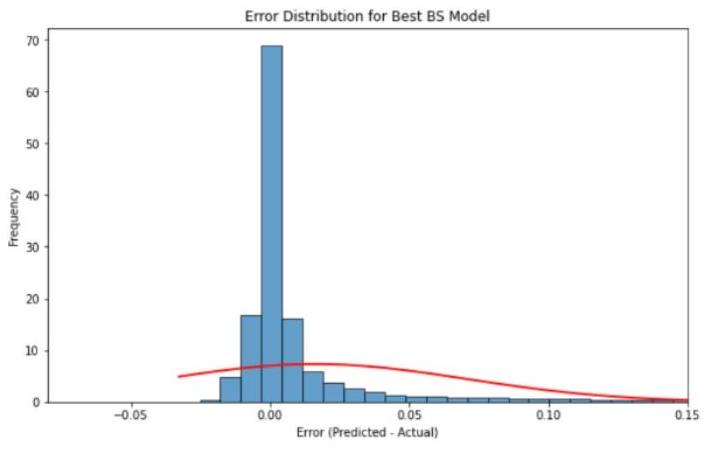}
\end{center}

\begin{itemize}
  \item Figure 9: Training \& Validation Loss Over Epochs for Best MLP Model
\end{itemize}

\begin{center}
\includegraphics[max width=\textwidth]{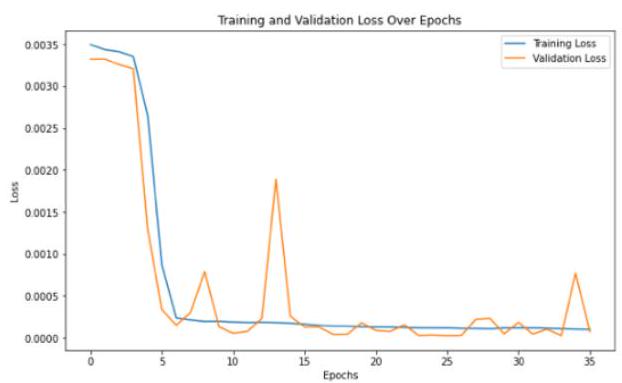}
\end{center}

\begin{itemize}
  \item Figure 10: Predicted vs Actual C/K for Best MLP Model
\end{itemize}

\begin{center}
\includegraphics[max width=\textwidth]{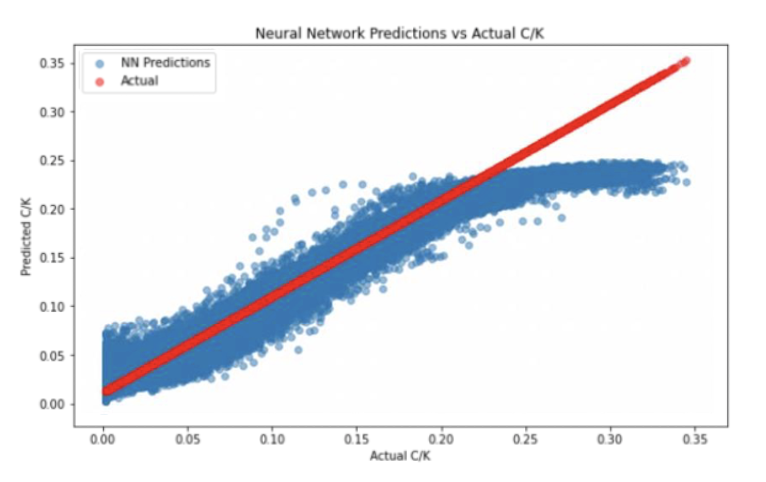}
\end{center}

\begin{itemize}
  \item Figure 11: Error Distribution for Best MLP Model
\end{itemize}

\begin{center}
\includegraphics[max width=\textwidth]{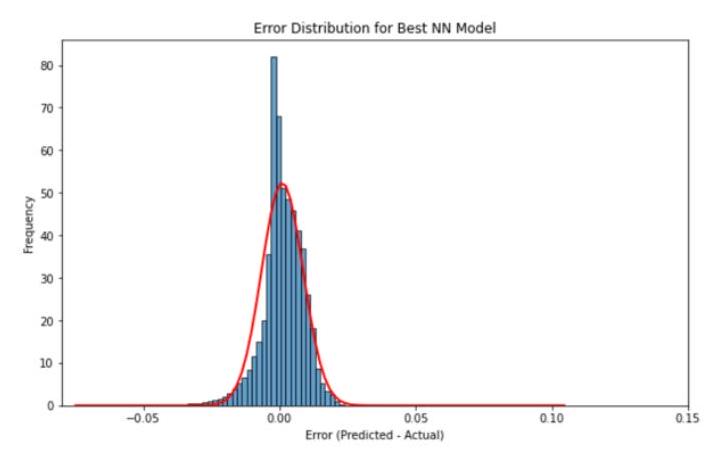}
\end{center}

\begin{itemize}
  \item Figure 12: Actual \& Best B-S Model C/K Predictions vs Time
\end{itemize}

\begin{center}
\includegraphics[max width=\textwidth]{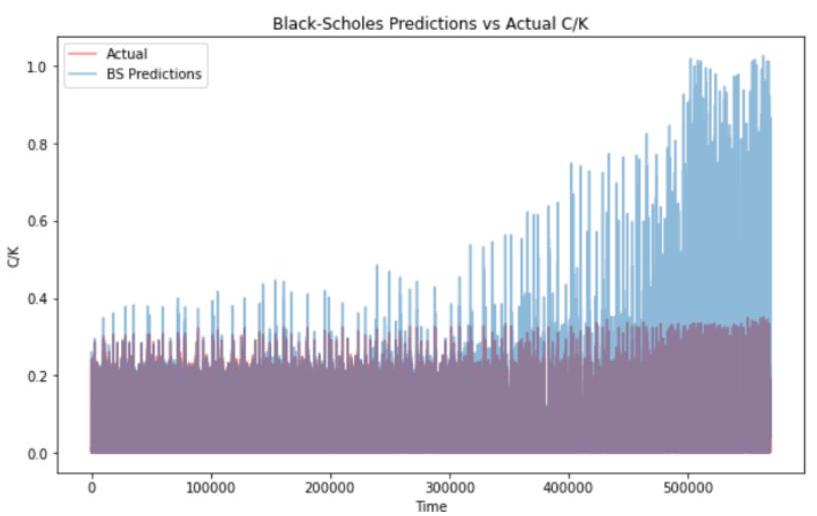}
\end{center}

\begin{itemize}
  \item Figure 13: Actual \& Best MLP Model C/K Predictions vs Time
\end{itemize}

\begin{center}
\includegraphics[max width=\textwidth]{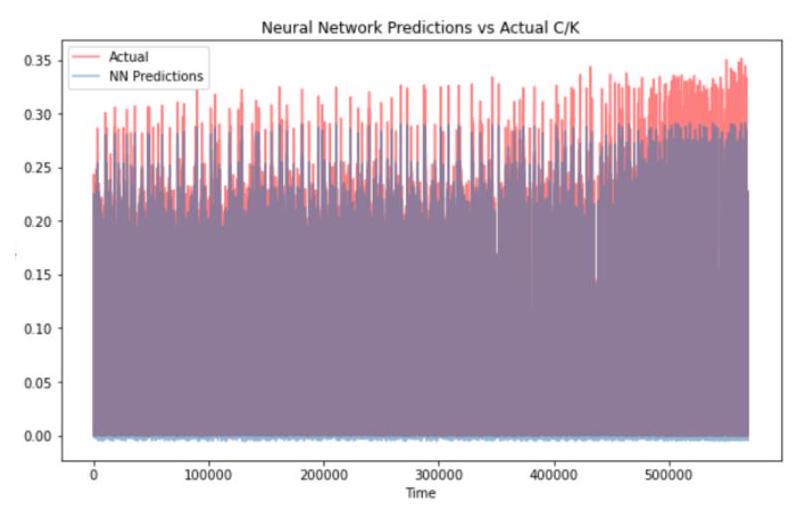}
\end{center}

\begin{itemize}
  \item Figure 14: Error Metrics for Best MLP \& B-S Models
\end{itemize}

\begin{center}
\begin{tabular}{|l|l|l|l|}
\hline
Error Metrics-> & MSE (6 d.p.) & RMSE (6 d.p.) & MAE (6 d.p.) \\
\hline
\begin{tabular}{l}
90d-vol. BS Model \\
(Best) \\
\end{tabular} & 0.003142 & 0.056056 & 0.018677 \\
\hline
Best MLP Model & 0.000056 & 0.007449 & 0.005483 \\
\hline
\end{tabular}
\end{center}

\begin{itemize}
  \item Figure 15: Activation Functions (ReLU, Swish, Tanh, GELU)
\end{itemize}

\begin{center}
\includegraphics[max width=\textwidth]{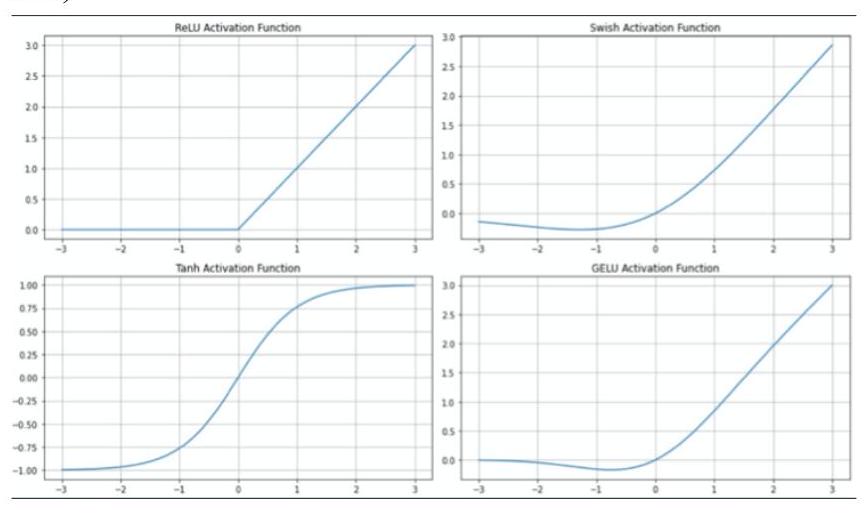}
\end{center}

\begin{itemize}
  \item Figure 16: Decision Tree for Prediction (\cite{Soni2020})
\end{itemize}

\begin{center}
\includegraphics[max width=\textwidth]{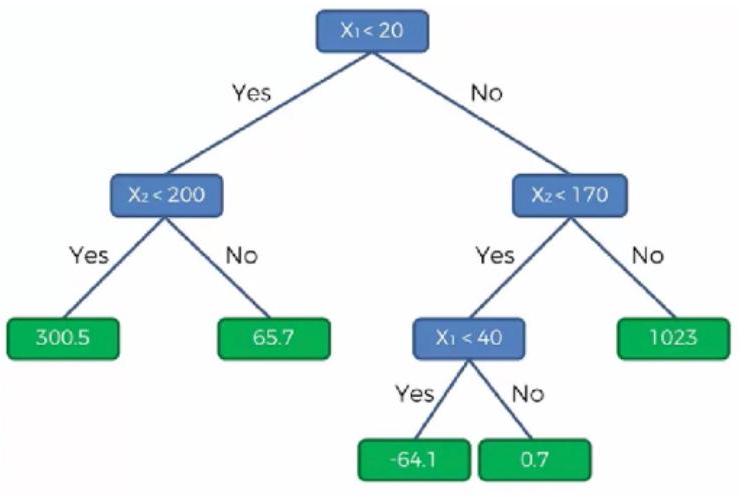}
\end{center}

\begin{itemize}
  \item Figure 17: XGBoost Prediction (\cite{Demajo2020})
\end{itemize}

\begin{center}
\includegraphics[max width=\textwidth]{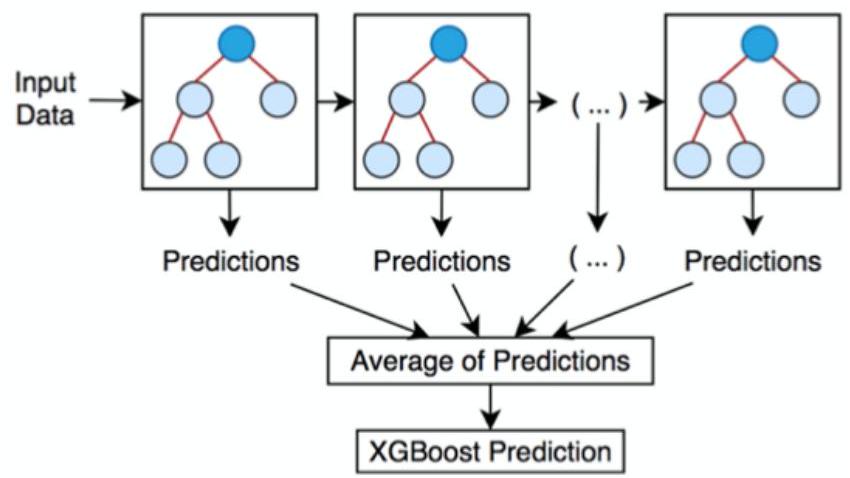}
\end{center}

\begin{itemize}
  \item Figure 18: Training \& Validation Loss Over Epochs for Best XGBoost Model
\end{itemize}

\begin{center}
\includegraphics[max width=\textwidth]{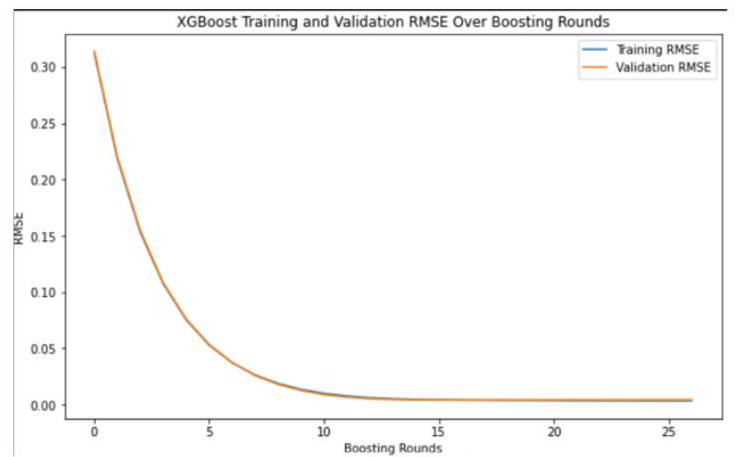}
\end{center}

\begin{itemize}
  \item Figure 19: Predicted vs Actual $\mathrm{C} / \mathrm{K}$ for Best XGBoost Model
\end{itemize}

\begin{center}
\includegraphics[max width=\textwidth]{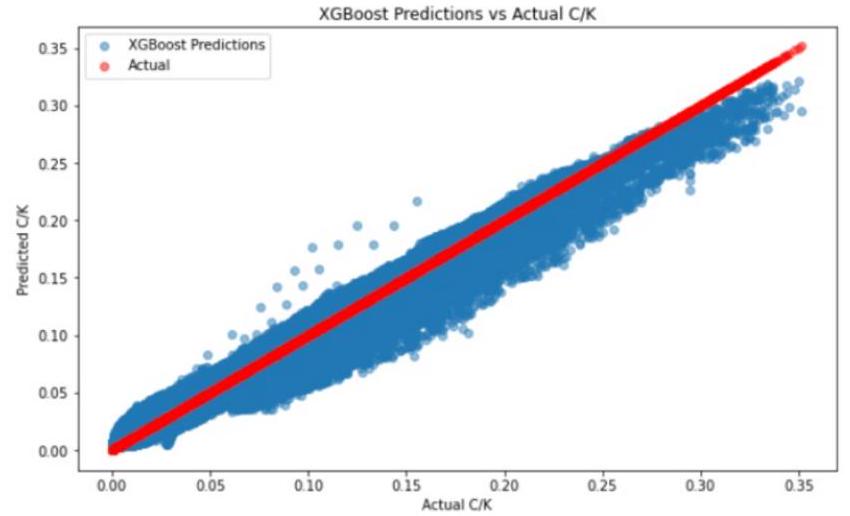}
\end{center}

\begin{itemize}
  \item Figure 20: Error Distribution for Best XGBoost Model
\end{itemize}

\begin{center}
\includegraphics[max width=\textwidth]{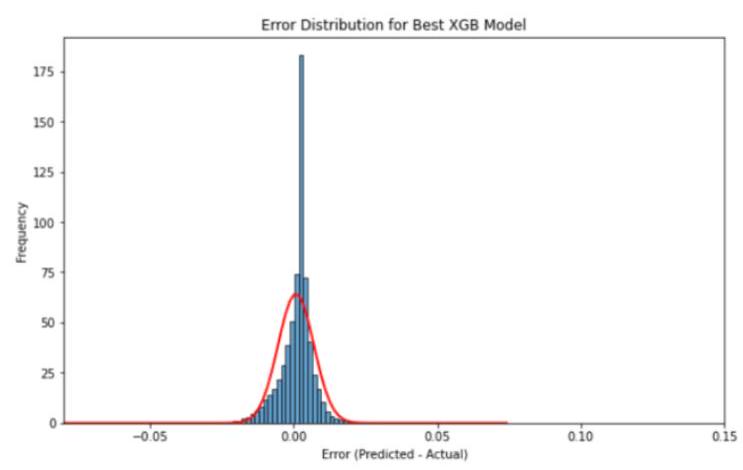}
\end{center}

\begin{itemize}
  \item Figure 21: Actual \& Best XGBoost Model C/K Predictions vs Time
\end{itemize}

\begin{center}
\includegraphics[max width=\textwidth]{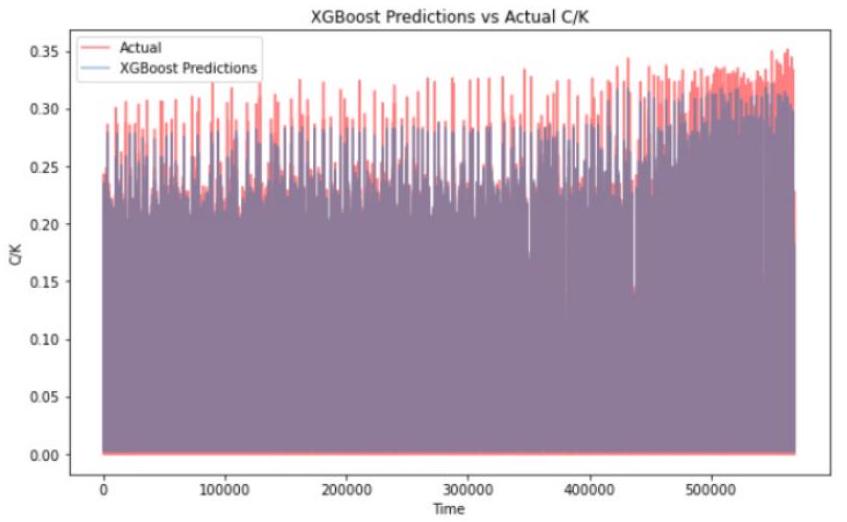}
\end{center}

\begin{itemize}
  \item Figure 22: Error Metrics for Best MLP, XGBoost, \& B-S Models
\end{itemize}

\begin{center}
\begin{tabular}{|l|l|l|l|}
\hline
Error Metrics-> & MSE (6 d.p.) & RMSE (6 d.p.) & MAE (6 d.p.) \\
\hline
90d-vol. BS Model (Best) & 0.003142 & 0.056056 & 0.018677 \\
\hline
Best MLP Model & 0.000056 & 0.007449 & 0.005483 \\
\hline
Best XGBoost Model & 0.000025 & 0.005021 & 0.003660 \\
\hline
\end{tabular}
\end{center}

\begin{itemize}
  \item Figure 23: Training \& Validation Loss Over Epochs for Best TDNN Model
\end{itemize}

\begin{center}
\includegraphics[max width=\textwidth]{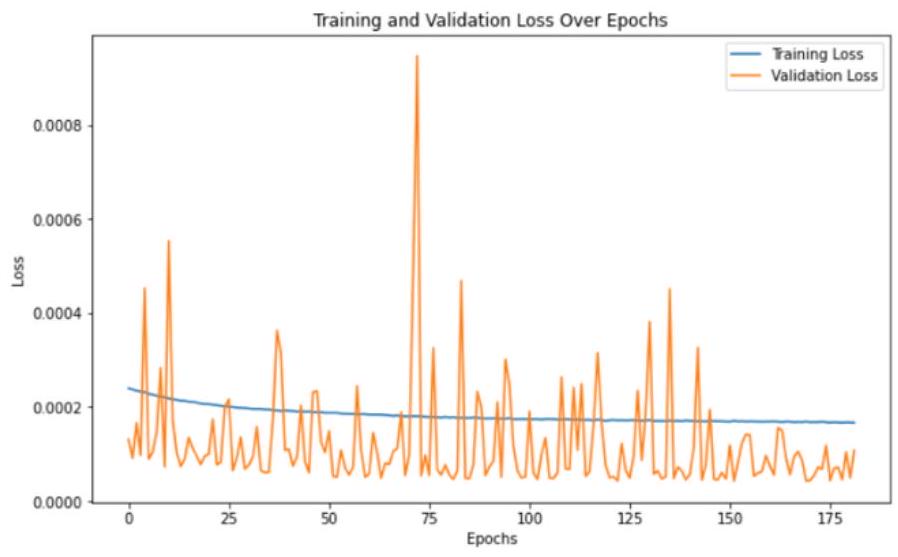}
\end{center}

\begin{itemize}
  \item Figure 24: Predicted vs Actual C/K for Best TDNN Model
\end{itemize}

\begin{center}
\includegraphics[max width=\textwidth]{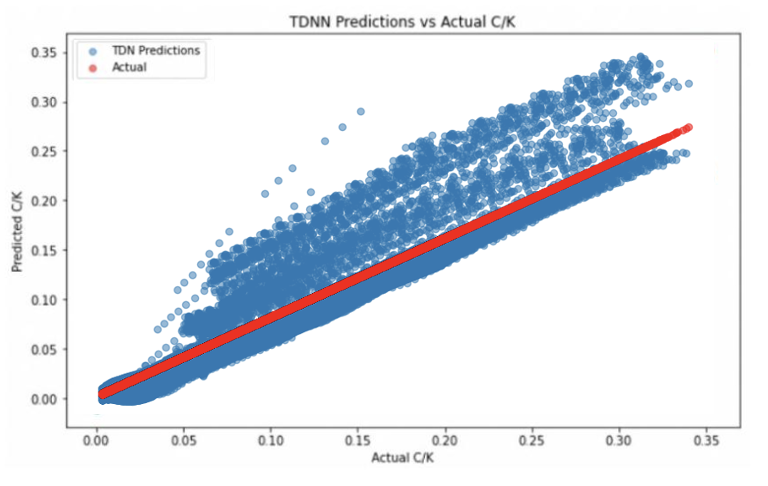}
\end{center}

\begin{itemize}
  \item Figure 25: Error Distribution for Best TDNN Model
\end{itemize}

\begin{center}
\includegraphics[max width=\textwidth]{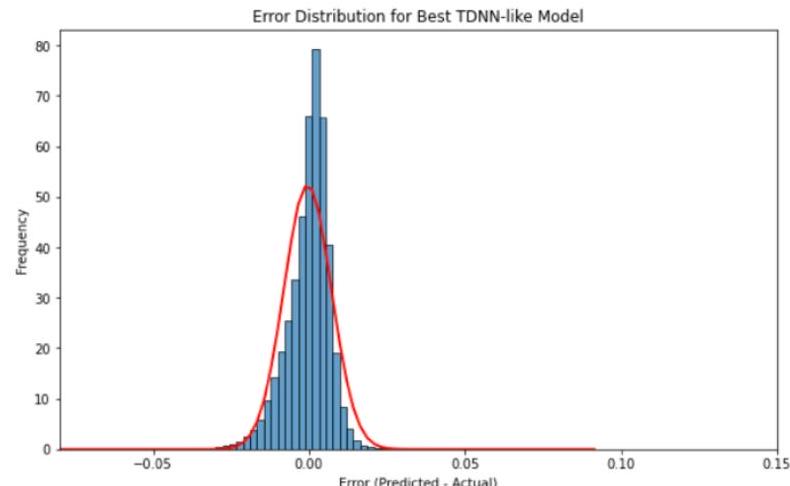}
\end{center}

\begin{itemize}
  \item Figure 26: Actual \& Best TDNN Model C/K Predictions vs Time
\end{itemize}

\begin{center}
\includegraphics[max width=\textwidth]{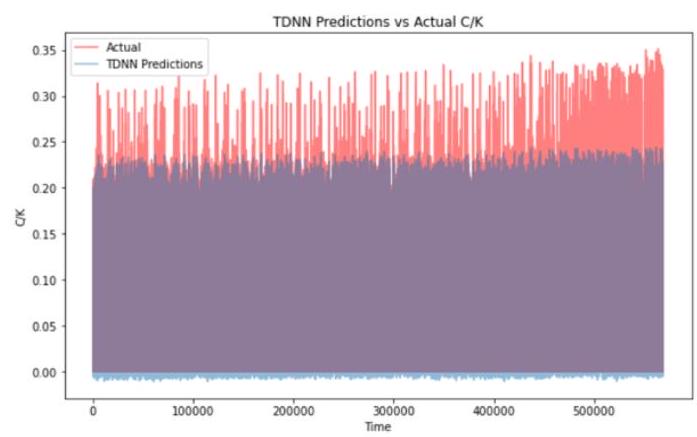}
\end{center}

\begin{itemize}
  \item Figure 27: Error Metrics by Model
\end{itemize}

\begin{center}
\begin{tabular}{|l|l|l|l|}
\hline
Error Metrics -> & MSE (6 d.p.) & RMSE (6 d.p.) & MAE (6 d.p.) \\
\hline
(1) Best RNN & 0.000021 & 0.004583 & 0.003301 \\
\hline
(2) Best XGBoost & 0.000025 & 0.005021 & 0.003660 \\
\hline
(3) Best KAN & 0.000039 & 0.006249 & 0.004425 \\
\hline
(4) Best MLP & 0.000056 & 0.007449 & 0.005483 \\
\hline
(5) Best TDNN & 0.000059 & 0.007682 & 0.005699 \\
\hline
(6) Best BS & 0.003142 & 0.056056 & 0.018677 \\
\hline
\end{tabular}
\end{center}

\begin{itemize}
    \item Figure 28: Total Percentages of Underpriced and Overpriced Options by Model
\end{itemize}

\begin{center}
\begin{tabular}{|l|l|l|l|l|l|l|}
\hline
\begin{tabular}{l}
Total Percentages of \\
Underpriced and \\
Overpriced Options by Model \\
\end{tabular} & BS & TDNN & MLP & KAN & XGBoost & RNN \\
\hline
Overpriced & $59.29 \%$ & $60.91 \%$ & $27.59 \%$ & $21.42 \%$ & $19.03 \%$ & $17.94 \%$ \\
\hline
Correctly Priced & $16.98 \%$ & $20.35 \%$ & $28.20 \%$ & $34.52 \%$ & $42.00 \%$ & $48.05 \%$ \\
\hline
Underpriced & $23.72 \%$ & $18.75 \%$ & $44.21 \%$ & $44.06 \%$ & $38.97 \%$ & $34.01 \%$ \\
\hline
\end{tabular}
\end{center}

\begin{itemize}
  \item Figure 29: Overpriced \& Underpriced Percentages by Ticker for Each Model
\end{itemize}

\begin{center}
\begin{tabular}{|c|c|c|c|}
\hline
\begin{tabular}{l}
 \\
Underpriced \\
Percentages by \\
Ticker for Each \\
Model \\
\end{tabular} & \begin{tabular}{l}
Underpriced \\
\% \\
\end{tabular} & \begin{tabular}{l}
Correctly \\
Priced \% \\
\end{tabular} & Overpriced \% \\
\hline
Best RNN: & -- & -- & -- \\
\hline
NDX & 32.18 & 45.96 & 21.86 \\
\hline
SPX & 36.02 & 50.44 & 13.54 \\
\hline
Best XGBoost: & -- & -- & -- \\
\hline
NDX & 30.23 & 40.50 & 29.27 \\
\hline
SPX & 39.01 & 44.65 & 16.34 \\
\hline
Best KAN: & -- & -- & -- \\
\hline
NDX & 47.02 & 33.59 & 19.39 \\
\hline
SPX & 41.06 & 35.00 & 23.94 \\
\hline
Best MLP: & -- & -- & -- \\
\hline
NDX & 45.06 & 28.03 & 26.91 \\
\hline
SPX & 43.43 & 32.21 & 24.36 \\
\hline
Best TDNN: & -- & -- & -- \\
\hline
NDX & 17.13 & 18.85 & 64.02 \\
\hline
SPX & 18.93 & 22.08 & 58.99 \\
\hline
Best B-S: & -- & -- & -- \\
\hline
NDX & 29.57 & 14.42 & 56.01 \\
\hline
SPX & 22.43 & 16.56 & 61.59 \\
\hline
\end{tabular}
\end{center}

\begin{itemize}
  \item Figure 30: Percentages of Overpriced \& Underpriced by Moneyness Category for Each Model
\end{itemize}

\begin{center}
\begin{tabular}{|c|c|c|c|}
\hline
\begin{tabular}{l}
Percentages of \\
Overpriced and \\
Underpriced \\
Options by \\
Moneyness \\
Category for Each \\
Model \\
\end{tabular} & \begin{tabular}{l}
Underpriced \\
\% \\
\end{tabular} & \begin{tabular}{l}
Correctly \\
Priced \% \\
\end{tabular} & Overpriced \% \\
\hline
Best RNN: & -- & -- & -- \\
\hline
ATM & 32.10 & 29.23 & 38.67 \\
\hline
ITM & 36.02 & 54.55 & 9.43 \\
\hline
OTM & 29.78 & 15.10 & 55.12 \\
\hline
Best XGBoost: & -- & -- & -- \\
\hline
ATM & 29.86 & 24.67 & 45.47 \\
\hline
ITM & 13.46 & 74.61 & 11.93 \\
\hline
OTM & 11.09 & 5.41 & 83.50 \\
\hline
Best KAN: & -- & -- & -- \\
\hline
ATM & 34.71 & 22.83 & 42.46 \\
\hline
ITM & 55.20 & 61.12 & 12.68 \\
\hline
OTM & 27.29 & 8.50 & 64.21 \\
\hline
Best MLP: & -- & -- & -- \\
\hline
ATM & 15.66 & 14.43 & 69.91 \\
\hline
ITM & 17.76 & 57.82 & 24.42 \\
\hline
OTM & 57.62 & 4.14 & 38.24 \\
\hline
Best TDNN: & -- & -- & -- \\
\hline
ATM & 34.71 & 20.83 & 44.46 \\
\hline
ITM & 27.29 & 61.12 & 11.59 \\
\hline
OTM & 31.32 & 4.91 & 63.77 \\
\hline
Best B-S: & -- & -- & -- \\
\hline
ATM & 31.07 & 10.32 & 58.61 \\
\hline
ITM & 16.02 & 55.58 & 28.40 \\
\hline
OTM & 22.68 & 0.69 & 76.63 \\
\hline
\end{tabular}
\end{center}

\begin{itemize}
  \item Figure 31: Actual \& Best RNN Model C/K Predictions  
\end{itemize}

\begin{center}
\includegraphics[max width=\textwidth]{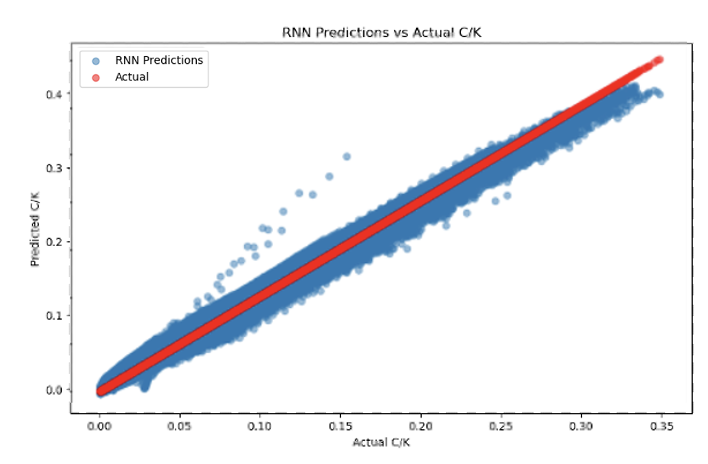}
\end{center}

\begin{itemize}
  \item Figure 32: Actual \& Best RNN Model C/K Predictions vs Time 
\end{itemize}

\begin{center}
\includegraphics[max width=\textwidth]{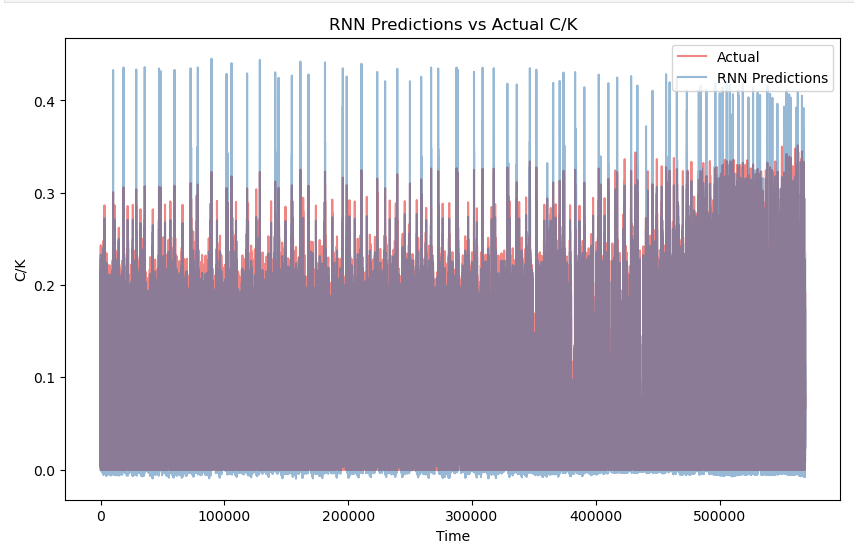}
\end{center}

\begin{itemize}
  \item Figure 33: Training \& Validation Loss Over Epochs for Best RNN Model
\end{itemize}

\begin{center}
\includegraphics[max width=\textwidth]{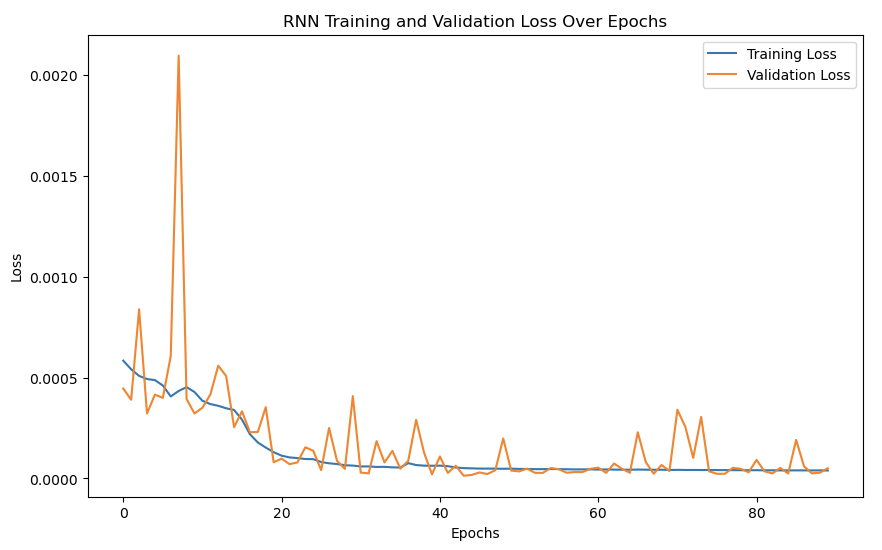}
\end{center}

\begin{itemize}
  \item Figure 34: Error Distribution for Best RNN Model
\end{itemize}

\begin{center}
\includegraphics[max width=\textwidth]{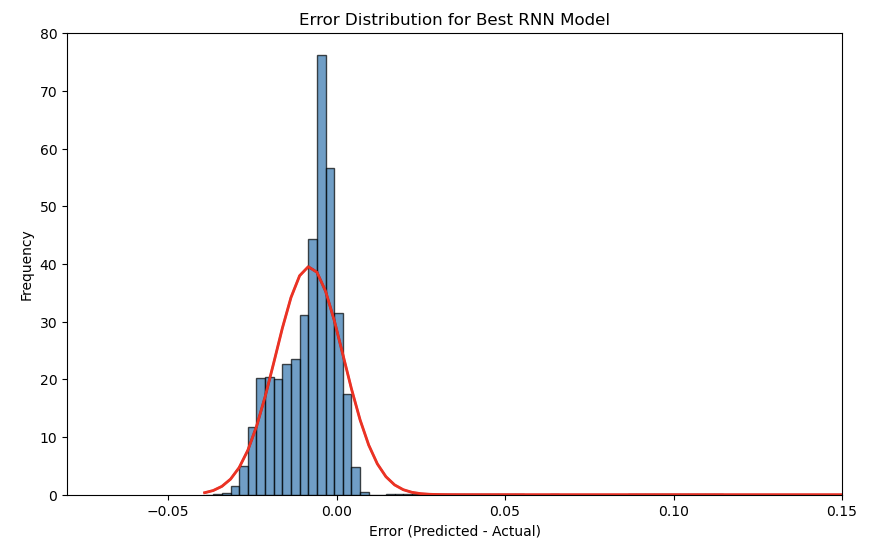}
\end{center}

\begin{itemize}
  \item Figure 35: Actual \& Best KAN Model C/K Predictions  
\end{itemize}

\begin{center}
\includegraphics[max width=\textwidth]{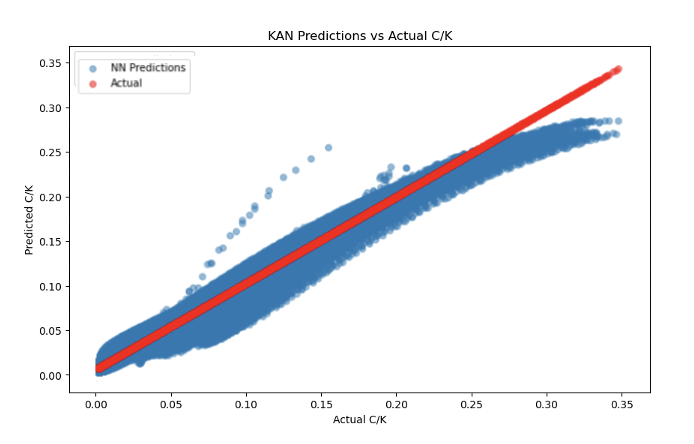}
\end{center}

\begin{itemize}
  \item Figure 36: Actual \& Best KAN Model C/K Predictions vs Time 
\end{itemize}

\begin{center}
\includegraphics[max width=\textwidth]{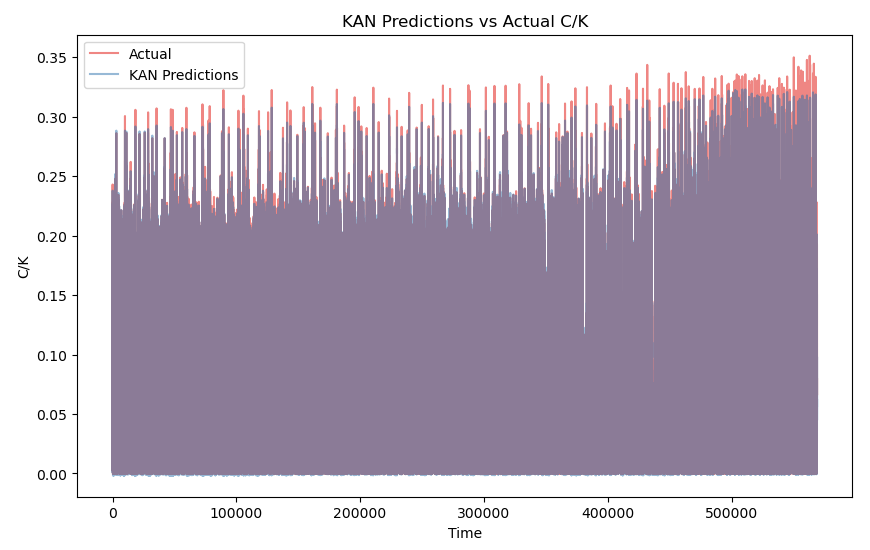}
\end{center}

\begin{itemize}
  \item Figure 37: Training \& Validation Loss Over Epochs for Best KAN Model
\end{itemize}

\begin{center}
\includegraphics[max width=\textwidth]{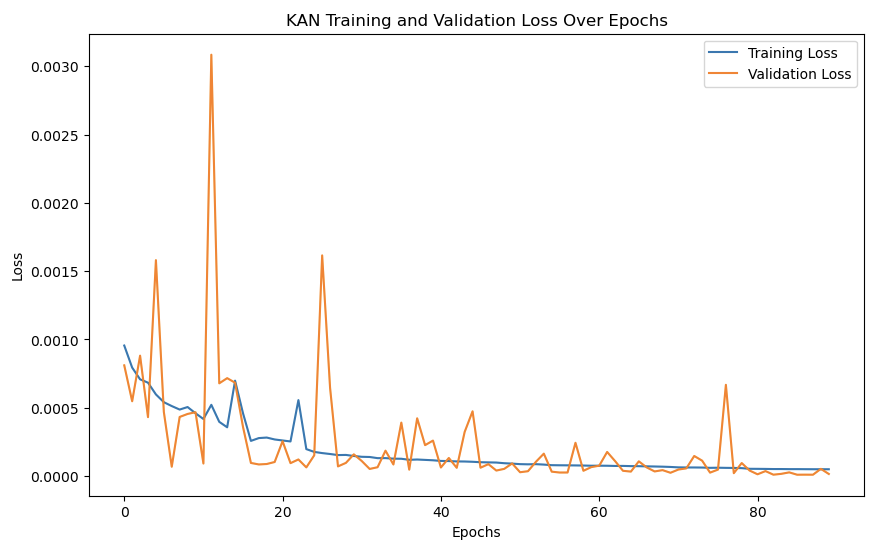}
\end{center}

\begin{itemize}
  \item Figure 38: Error Distribution for Best KAN Model
\end{itemize}

\begin{center}
\includegraphics[max width=1.1\textwidth]{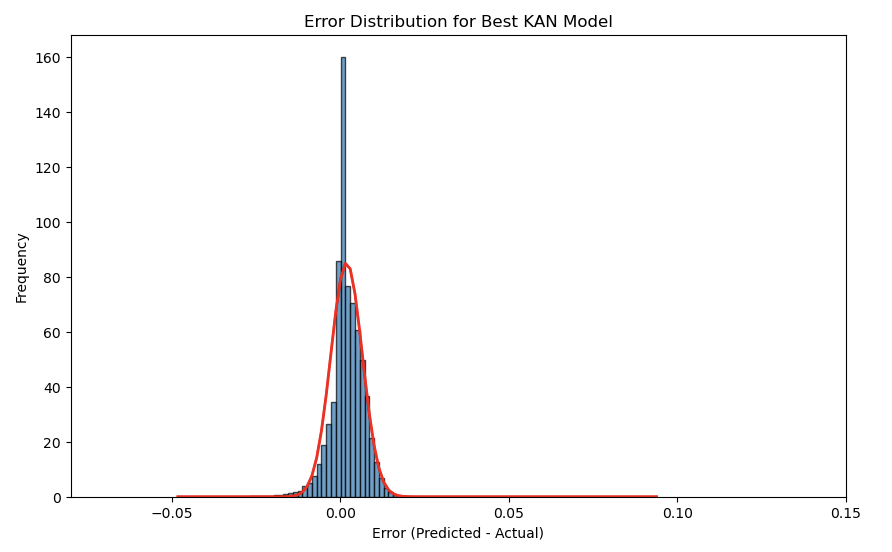}
\end{center}

\begin{itemize}
  \item Figure 39: First 7 Orthogonal Polynomials for Besel, Laguerre, Legendre, and Chebyshev of 2nd Kind 
\end{itemize}

\begin{center}
\includegraphics[max width=\textwidth]{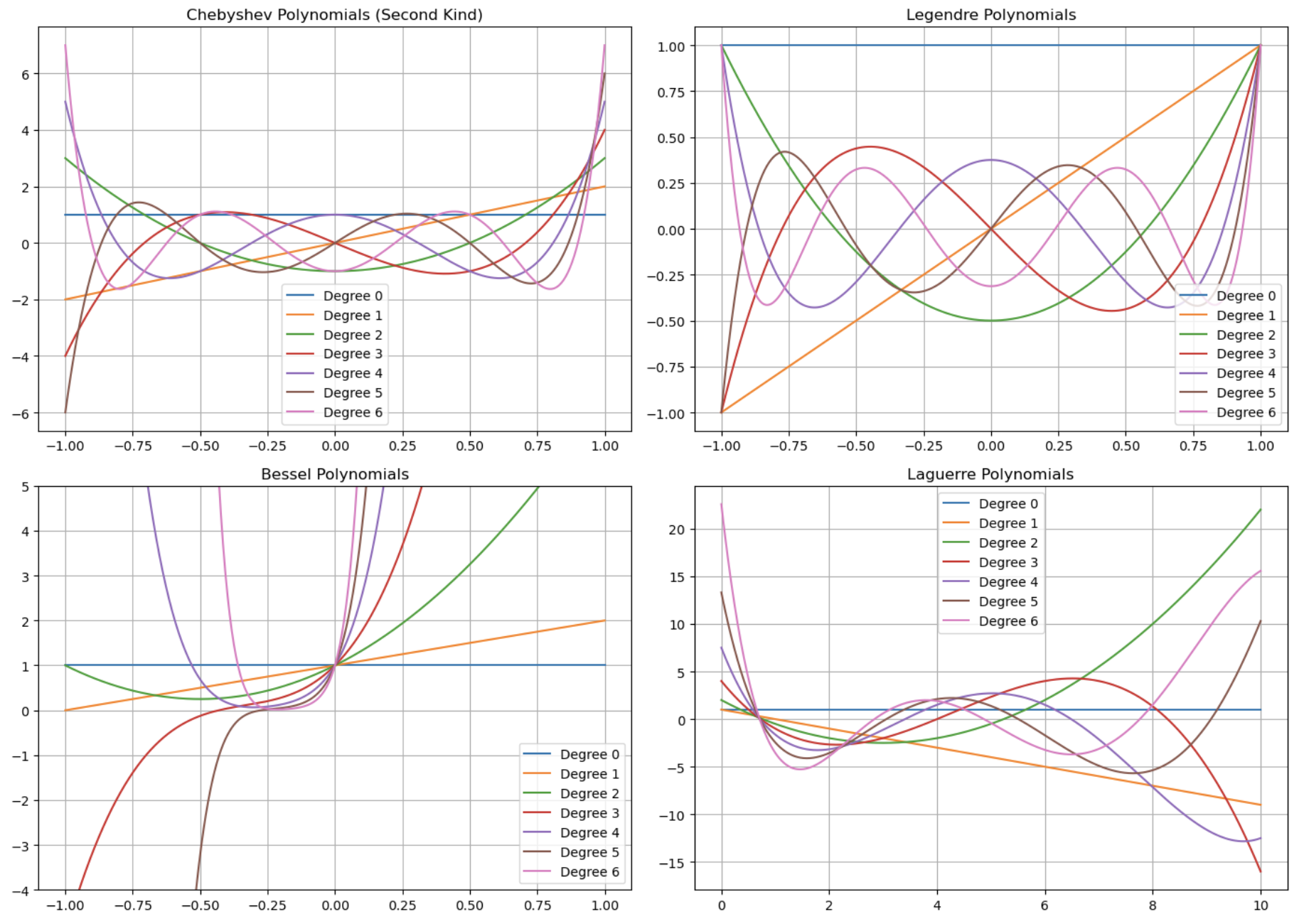}
\end{center}

\begin{itemize}
  \item Figure 40: (Top) All Options contracts by K vs T ; (Bottom) SPX and NDX price over the same period. 
\end{itemize}

\begin{center}
\includegraphics[max width=\textwidth]{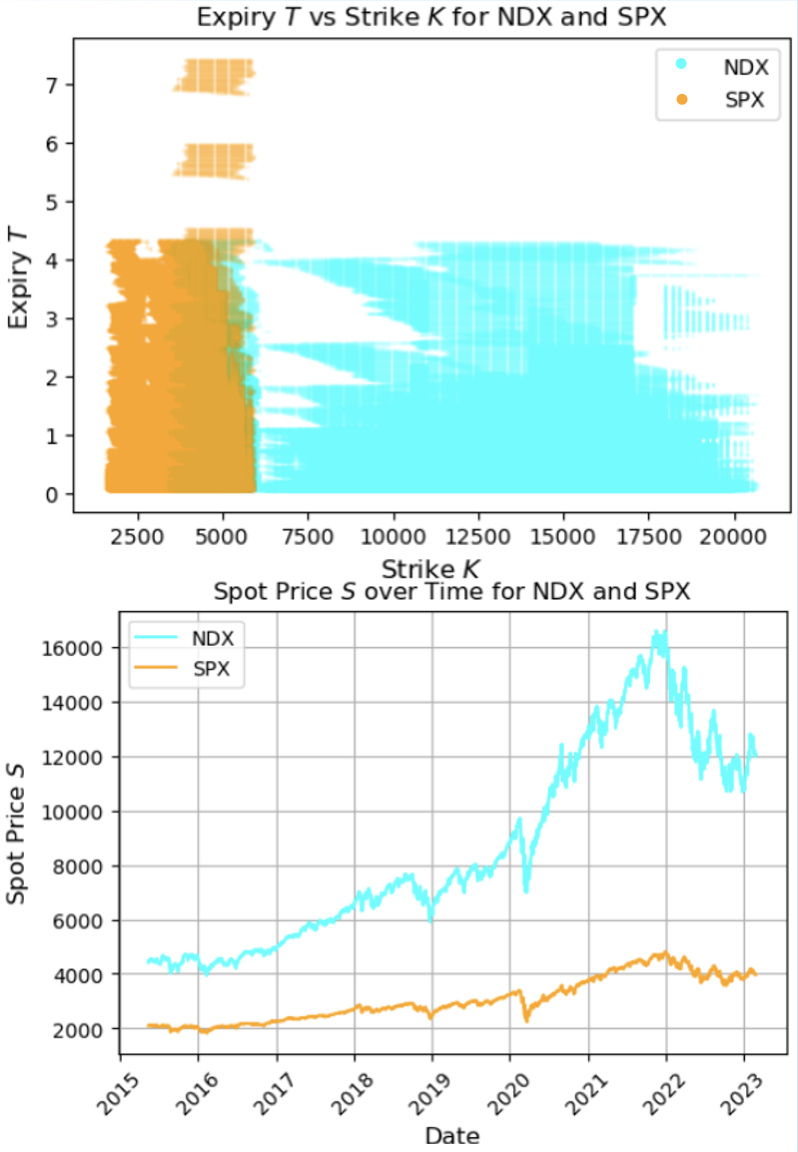}
\end{center}

\end{document}